\documentclass[twocolumn,aps,pre,groupedaddress]{revtex4-1}

\usepackage{amsmath}
\usepackage{graphicx}
\usepackage{amsmath}
\usepackage{graphicx}
\usepackage{subfig}
\usepackage{epstopdf}
\usepackage{amssymb}
\usepackage{multirow}
\usepackage[english]{babel}
\usepackage{natbib}

\begin{document}
\title{Non-parametric causal inference for bivariate time series}
\author{James M. McCracken}
\email{jmccrac1@masonlive.gmu.edu}
\affiliation{Department of Physics and Astronomy \\ George Mason University \\ 4400 University Drive MS 3F3, Fairfax, VA 22030-4444}
\author{Robert S. Weigel}
\email{rweigel@gmu.edu}
\affiliation{Department of Physics and Astronomy \\ George Mason University \\ 4400 University Drive MS 3F3, Fairfax, VA 22030-4444}
\date{\today}

\begin{abstract}
We introduce new quantities for exploratory causal inference between bivariate time series.  The quantities, called penchants and leanings, are computationally straightforward to apply, follow directly from assumptions of probabilistic causality, do not depend on any assumed models for the time series generating process, and do not rely on any embedding procedures; these features may provide a clearer interpretation of the results than those from existing time series causality tools.  The penchant and leaning are computed based on a structured method for computing probabilities.
\end{abstract}

\maketitle

\section{Introduction}
Many scientific disciplines rely on observational data from systems in which it is difficult or impossible to implement controlled experiments or to control interventions. For example, there is no current technology that can control the interaction between the solar wind and the magnetic field measured at the surface of Earth, so space weather studies rely on data collected without performing controlled experiments.  As a result, causal inference with observational data sets from such systems is difficult and the need to identify causal relationships given the weakness of correlation in doing so has lead to the development of several different time series causality tools \cite{Schreiber2000,granger1969,Rogosa1980,Pearl2000,kleinberg2012}.

Causal inference in time series involves finding ``driving'' relationships between different time series signals.  Showing the existence, rather than the exact nature, of the driving relationship between the signals is often the primary goal.  Thus, words like ``driving'', ``causality'', and related terms typically do not have straightforward analogs to the same terms used in other fields \cite{Granger1980,liu2012,Roberts1985}, e.g.\, theoretical quantum (e.g., \cite{Peres1995}) or classical mechanics (e.g., \cite{Bunge1979}).

The development and study of causal inference techniques is often called {\em time series causality}.  Most techniques fall into four broad categories related to either transfer entropy \cite{Schreiber2000}, Granger causality \cite{granger1969}, state space reconstruction (SSR) \cite{Sugihara2012}, or lagged cross-correlation \cite{box2013,pascual2014}.  These techniques have found application in a wide range of fields including neuroscience (e.g.,\ \cite{Kaminski2001}), economics (e.g.,\ \cite{dufour1998,dufour2006}), and climatology (e.g.,\ \cite{mosedale2006}).  

In this article, we introduce a time series causality technique derived directly from the definition of probabilistic causality \cite{Suppes1970}.  The technique is applied to synthetic and  empirical bivariate time series data sets with known, or intuitive, causal relationships.  We discuss the strengths and weaknesses of the technique and demonstrate how it may be useful for causal inference with empirical data from systems in which it is difficult or impossible to implement controlled experiments or to control interventions.  

\section{Causal Penchant}
We define the {\em causal penchant} $\rho_{EC}\in\left[1,-1\right]$ as
\begin{equation}
\label{eq:pen}
\rho_{EC} := P\left(E|C\right) - P\left(E|\bar{C}\right).
\end{equation}

The motivation for this expression is in the straightforward interpretation of $\rho_{EC}$ as a causal indicator \footnote{Eqn.\ \ref{eq:pen} may appear related to the {\em weight of evidence} \cite{Good1984}.  The two quantities, however, are fundamentally different, both mathematically and conceptually.  A requirement of weight of evidence is that the posterior probability should be recoverable provided probability of the evidence alone; i.e., $P(C|E)$ should be derivable from $\rho_{EC}$ and $P(E)$ if $\rho_{EC}$ can be considered weight of evidence as Good describes (see Eqn.\ (2) of \cite{Good1984}).  This requirement is not met by the penchant, as the probability of the cause, $P(C)$, is also required to derive the posterior.  The intended scope of the penchant is also much more restricted than that of the weight of evidence.  Most of the examples presented by Good in \cite{Good1984} cannot be addressed directly with any times series causality tools.  Eqn.\ \ref{eq:pen} is, however, identical in form to the {\em Eells measure of causal strength} (also called the {\em probability contrast}) \cite{Illari2011}.  Eqn.\ \ref{eq:pen} will not be used as a measure of causal strength in this work.  Rather, it will be used as a measure indicative of driving without any concern for notions of the ``strength'' of such driving.  As such, we will use the term {\em penchant} to avoid confusion.  Also see \cite{kleinberg2012} for a discussion of similar expressions.}; i.e.,\ if $C$ causes (or {\em drives}) $E$, then $\rho_{EC} > 0$, and if $\rho_{EC} \le 0$, then the direction of causal influence is undetermined.  If effect $E$ is assumed to be measured in one time series and the cause $C$ is assumed to be measured in a different time series, then the direction of causal influence can be determined by comparing various penchants when each time series is assigned to be the cause $C$.  

Eqn.\ \ref{eq:pen} can be rewritten using Bayes' theorem 
\begin{equation}
\label{eq:bayes}
P(E|C) = P(C|E)\frac{P(E)}{P(C)}
\end{equation}
and the definitions of probability complements
\begin{equation}
\label{eq:comp1}
P(\bar{C}) = 1-P(C)
\end{equation}
\begin{equation}
\label{eq:comp2}
P(\bar{C}|E) = 1-P(C|E).  
\end{equation}
Using Eqn.\ \ref{eq:comp2} with Eqn.\ \ref{eq:bayes} gives 
\begin{eqnarray*}
P(\bar{C}|E) &=& 1-P(E|C)\frac{P(C)}{P(E)}
\end{eqnarray*}
Inserting this into Eq.\ \ref{eq:bayes} written in terms of $\bar{C}$\;\;,
\begin{eqnarray*}
P(E|\bar{C}) &=& P(\bar{C}|E)\frac{P(E)}{P(\bar{C})}\\
\end{eqnarray*}
yields an alternative form of the second term in Eqn.\ \ref{eq:pen}
\begin{eqnarray*}
P(E|\bar{C}) &=&\left(1-P(E|C)\frac{P(C)}{P(E)}\right)\frac{P(E)}{1-P(C)}\;\;,
\end{eqnarray*}
This expression gives a penchant that requires only a single conditional probability estimate:
\begin{equation}
\label{eq:pencal}
\rho_{EC} = P(E|C)\left(1+\frac{P(C)}{1-P(C)}\right)-\frac{P(E)}{1-P(C)}\;\;.
\end{equation}

The penchant is non-parametric in the sense that there is no assumed functional relationship between the time series being investigated.  This form of the penchant will be used along with a structured method for counting and notions from probabilistic causality to infer which time series in a given pair might be seen as ``driving'' the other.  Our motivation for the penchant is the need for a time series causality quantity that is easily computed.  

For the calculations in the following sections, the penchant is not defined if $P(C)$ or $P(\bar{C})$ are zero (because the conditionals in Eqn.\ \ref{eq:pen} would be undefined).  Thus, the penchant is not defined if $P(C)=0$ or if $P(C)=1$.  The former condition corresponds to an inability to determine causal influence between two time series when a cause does not appear in one of the series; the latter condition is interpreted as an inability to determine causal influence between two time series if one is constant.  The use of Bayes' theorem in the derivation of Eqn.\ \ref{eq:pencal} implies that the penchant is not defined if $P(E)$ or $P(\bar{E})$ are zero.

The method given in this work uses no {\em a priori} assignment of ``cause'' or ``effect'' to a given time series pair when using penchants for causal inference.  So, operationally, the constraints on $P(C)$ and $P(E)$ only mean that the penchant is undefined between pairs of time series where one series is constant. 

The penchant definition includes $P(E|\bar{C})$, which is the probability of an assumed effect occurring given a absence of the assumed cause.  It has been argued that causality determination requires an intervention, and the absence of an assumed cause is unobservable, which implies the occurrence probability of the assumed effect should be conditioned on performing or not performing an action rather than on the presence or lack of an assumed cause \cite{Pearl2000}.  Causal relations have been described as ``a relation among events'' \cite{Bunge1979}, again implying the absence of an assumed cause cannot be used to identify causal relationships.  These issues have been a part of probabilistic definitions of causality at least since the 1960s \cite{Suppes1970}, and we do not attempt to solve them in this article.  We circumvent these philosophical issues by using an expression that removes any conditioning on the absence of an assumed cause and the condition that the penchant is undefined when $P(C)=0$, $P(C)=1$, $P(E)=0$, or $P(E)=1$. 

Although Eqn.\ \ref{eq:pencal} circumvents the issue of $P(E|\bar{C})$ being unobservable, it does not account for confounding.  The assumption that $P(C)$ can be estimated from a scalar time series may be seen as an oversimplification of the dynamics.  That is, it may be seen as an assumption that the assumed effect is only caused by the assumed cause.  In this case, the penchant may be better interpreted as an indication of predictability rather than causality (similar to arguments made regarding Granger causality \cite{Sugihara2012}).  This issue will not be addressed in this article;  we emphasize, however, that we use terms such as cause, effect, causal inference, and related terms to specifically refer to the penchant and leaning quantities.  

In this article, we seek to determine if the penchant is a useful quantity for the identification of causality relationships between time series.  Our goal is identify the usefulness of the penchant (and leaning introduced below) for {\em exploratory causal inference}, i.e., inference intended to determine if (and what) causal structure may be present in a time series pair but not to {\em prove} or {\em confirm} such structure.  There are scenarios in which any time series causality tool such as the penchant may incorrectly assign causal structure or may incorrectly not assign causal structure \cite{Illari2014}.  Furthermore, proof of causal relationships is often considered impossible with data alone \cite{Pearl2000,Illari2014,Rubin2015}.  The goal of this work is to draw as much information as possible from the given data to, e.g., guide the design of future experiments.

\section{Causal Leaning}
\label{sec:lean}
Consider the assignment of $\mathbf{X}$ as the cause, $C$, and $\mathbf{Y}$ as the effect, $E$.  If $\rho_{EC}>0$, then the probability that $\mathbf{X}$ drives $\mathbf{Y}$ is higher than the probability that it does not, which is stated more succinctly as $\mathbf{X}$ has a penchant to drive $\mathbf{Y}$ or $\mathbf{X}\xrightarrow{pen}\mathbf{Y}$.  

It is possible, however, that the penchant could be positive when $\mathbf{X}$ is assumed as the effect and $\mathbf{Y}$ is assumed as the cause.  (An example of this is given in Section \ref{sec:meanlean}.) The {\em leaning} addresses this via
\begin{equation}
\label{eq:leaning}
\lambda_{EC} := \rho_{EC} - \rho_{CE}
\end{equation}
for which $\lambda_{EC}\in\left[-2,2\right]$. A positive leaning implies the assumed cause $C$ drives the assumed effect $E$ more than the assumed effect drives the assumed cause, a negative leaning implies the effect $E$ drives the assumed cause $C$ more than the assumed cause drives the assumed effect, and a zero leaning yields no causal information.  

The possible outcomes are notated as
\begin{flalign*}
&\lambda_{EC}>0 \quad\{C,E\} = \{\mathbf{X},\mathbf{Y}\}\Rightarrow\mathbf{X}\xrightarrow{lean}\mathbf{Y}&\\
&\lambda_{EC}<0\quad\{C,E\} = \{\mathbf{X},\mathbf{Y}\}\Rightarrow\mathbf{Y}\xrightarrow{lean}\mathbf{X}&\\
&\lambda_{EC}=0\quad\{C,E\} = \{\mathbf{X},\mathbf{Y}\}\Rightarrow \mbox{no conclusion}&
\end{flalign*}
with $\{C,E\} = \{\mathbf{A},\mathbf{B}\}$ meaning $\mathbf{A}$ is the assumed cause and $\mathbf{B}$ as the assumed effect.

If $\lambda_{EC}>0$ with $\mathbf{X}$ as the assumed cause and $\mathbf{Y}$ as the assumed effect, then $\mathbf{X}$ has a larger penchant to drive $\mathbf{Y}$ than $\mathbf{Y}$ does to drive $\mathbf{X}$.  That is, $\lambda_{EC}>0$ implies that the difference between the probability that $\mathbf{X}$ drives $\mathbf{Y}$ and the probability that it does not is higher than the difference between the probability that $\mathbf{Y}$ drives $\mathbf{X}$ and the probability that it does not.  

The leaning is a function of four probabilities, $P(C)$, $P(E)$, $P(C|E)$, and $P(E|C)$.  The usefulness of the leaning for causal inference will depend on an effective method for estimating these probabilities from times series and a more specific definition of the cause-effect assignment within the time series pair.  An operational definition of $C$ and $E$ will need to be drawn directly from the time series data if the leaning is to be useful for causal inference.  Such assignments, however, may be difficult to develop and may be considered arbitrary without some underlying theoretical support.  For example, if the cause is $x_{t-1}$ and the effect is $y_{t}$, then it may be considered unreasonable to provide a causal interpretation of the leaning without theoretical support that $\mathbf{X}$ may be expected to drive $\mathbf{Y}$ on the time scale of $\Delta t=1$.  This issue is, however, precisely one of the reasons for divorcing the causal inference proposed in this work (i.e., exploratory causal inference) from traditional ideas of causality, as was explained in the second paragraph of the introduction.  Statistical tools are associational, and cannot be given formal causal interpretation without the use of assumptions and outside theories (see \cite{Illari2014} for an in-depth discussion of these ideas).  In practice, many different potential cause-effect assignments may be used to calculate different leanings, which may then be compared as part of the causal analysis of the data.

In this article, the probabilities required for the leaning calculation will be estimated from the data straightforwardly through a counting/binning procedure, and the cause-effect assignment may be varied but will always use a simple lag structure to avoid unnecessarily complex computations.  The process of estimating probabilities from time series data to draw causal inferences can be subtle; see the work of Schreiber et al.\ for a discussion of this issue in the context of transfer entropies \cite{kaiser2002}.  

\section{Motivating Example}
Consider a time series pair $\{\mathbf{X},\mathbf{Y}\}$ with
\begin{eqnarray*}
\label{eq:motex}
\mathbf{X} &=& \{x_t\; | \; t=0,1,2,\ldots,9 \}\\
&=& \left\{0,0,1,0,0,1,0,0,1,0\right\}\\
\mathbf{Y} &=& \{y_t\; | \; t=0,1,2,\ldots,9\}\\
&=& \left\{0,0,0,1,0,0,1,0,0,1\right\}.
\end{eqnarray*}

Because $y_t=x_{t-1}$, one may conclude that $\mathbf{X}$ drives $\mathbf{Y}$.  However, to show this result using a leaning calculation requires first a calculation using the cause-effect assignment $\{C,E\}=\{\mathbf{X},\mathbf{Y}\}$. For consistency with the intuitive definition of causality, we require that a cause must precede an effect.  It follows that a natural assignment may be $\{C,E\}=\{x_{t-l},y_t\}$ for $1 \leq l < t \leq 9$.  This cause-effect assignment will be referred to as the $l$-standard assignment.

The cause-effect assignment is an assignment of a given structure or feature of the data in one time series as the ``cause'' and another structure or feature of the data in the other time series as the ``effect''.  For example, in the $l$-standard cause-effect assignment, the cause is the lag $l$ time step in one time series and the effect is the current time step in the other.  The leaning compares the symmetric application of these cause-effect definitions to the time series pair.  So, for the above example of $\{C,E\}=\{x_{t-l},y_t\}$, the first penchant will be calculated using $\{C,E\}=\{x_{t-l},y_t\}$ and the second will be calculated using $\{C,E\}=\{y_{t-l},x_t\}$.  The second penchant is not the direct interchange of $C\Leftrightarrow E$ from the first penchant because such an interchange would violate the assumption that a cause must precede an effect.  For example, if the first penchant in the leaning calculation is calculated using $\{C,E\}=\{x_{t-l},y_t\}$, then the second penchant is not calculated using $\{C,E\}=\{y_t,x_{t-l}\}$ because the definition of the effect, $x_{t-l}$, precedes the definition of the cause, $y_t$.

\subsection{Defining penchants}
Given $\{\mathbf{X},\mathbf{Y}\}$, one possible penchant (i.e., Eqn.\ \ref{eq:pencal}) that can be defined using the 1-standard assignment is
\begin{eqnarray*}
\rho_{y_{t}=1,x_{t-1}=1} &=& \kappa \left(1+\frac{P\left(x_{t-1} = 1\right)}{1-P\left(x_{t-1} = 1\right)}\right)\\
& & -\frac{P\left(y_{t} = 1\right)}{1-P\left(x_{t-1} = 1\right)}\;\;,
\end{eqnarray*}
with $\kappa = P\left( y_t = 1 | x_{t-1} = 1\right)$.  Another penchant defined using this assignment is $\rho_{y_t=0,x_{t-1}=0}$ with  $\kappa = P\left( y_t = 0 | x_{t-1} = 0\right)$.  These two penchants are called {\em observed} penchants because they correspond to conditions that were found in the measurements.  

Two other penchants have $\kappa = P\left( y_t = 0 | x_{t-1} = 1\right)$ and $\kappa = P\left( y_t = 1 | x_{t-1} = 0\right)$.  These penchants are associated with unobserved conditions.  Based on the values for these two penchants, $\kappa=0\Rightarrow \rho_{y_{t}x_{t-1}} < 0$, which is consistent with the claim that the effect, $y_t = 0$ or $1$ is not caused by the postulated cause, $x_{t-1} = 1$ or $0$, respectively.  


\subsection{Computing penchants}
The probabilities in the penchant calculations can be estimated from time series using counts, e.g.,\
$$
P\left( y_t = 1 | x_{t-1} = 1\right) = \frac{n_{EC}}{n_C} = \frac{3}{3} = 1\;\;,
$$
where $n_{EC}$ is the number of times $y_t=1$ and $x_{t-1}=1$ appears in $\{\mathbf{X},\mathbf{Y}\}$, and $n_{C}$ is the number of times the assumed cause, $x_{t-1}=1$, has appeared in $\{\mathbf{X},\mathbf{Y}\}$.  

Estimating the other two probabilities in this penchant calculation using frequency counts from $\{\mathbf{X},\mathbf{Y}\}$ requires accounting for the assumption that the cause must precede the effect by shifting $\mathbf{X}$ and $\mathbf{Y}$ into $\tilde{\mathbf{X}}$ and $\tilde{\mathbf{Y}}$ such that, for any given $t$, $\tilde{\mathbf{x}}_t$ precedes $\tilde{\mathbf{y}}_t$.

For this example, the shifted sequences are
\begin{eqnarray*}
\tilde{\mathbf{X}} &=& \left\{0,0,1,0,0,1,0,0,1\right\}\\
\tilde{\mathbf{Y}} &=& \left\{0,0,1,0,0,1,0,0,1\right\}
\end{eqnarray*}
which are both shorter than there counterparts above by a single value because the penchants are being calculated using the 1-standard cause-effect assignment. It follows that $\tilde{x}_t = x_{t-1}$ and $\tilde{y}_t=y_t$.  

The probabilities are then
\begin{equation}
P\left( y_t = 1\right) = \frac{n_E}{L} = \frac{3}{9}
\end{equation}
and
\begin{equation}
P\left( x_{t-1} = 1\right) = \frac{n_C}{L} = \frac{3}{9}\;\;,
\end{equation}
where $n_C$ is the number of times $\tilde{x}_t = 1$, $n_E$ is the number of times $\tilde{y}_t = 1$, and $L$ is the (``library'') length of $\tilde{\mathbf{X}}$ and $\tilde{\mathbf{Y}}$.  

\subsection{Mean observed leaning}
\label{sec:meanlean}

The two observed penchants in this example under the assumption that $\mathbf{X}$ causes $\mathbf{Y}$ (with $l=1$) are
\begin{equation}
\label{eqn:rhoex1}
\rho_{y_t=1,x_{t-1}=1}=1
\end{equation}
and
\begin{equation*}
\rho_{y_t=0,x_{t-1}=0}=1\;\;.
\end{equation*}

The observed penchants when $\mathbf{Y}$ is assumed to cause $\mathbf{X}$ are
\begin{eqnarray*}
\rho_{x_t=1,y_{t-1}=0} &=& \frac{3}{7}\;\;,\\
\rho_{x_t=0,y_{t-1}=1} &=& \frac{3}{7}\;\;,
\end{eqnarray*}
and
\begin{equation*}
\rho_{x_t=0,y_{t-1}=0}=-\frac{3}{7}\;\;.
\end{equation*}

The {\em mean observed penchant} is the algebraic mean of the observed penchants,  For $\mathbf{X}$ causes $\mathbf{Y}$, it is
\begin{eqnarray*}
\langle \rho_{y_t,x_{t-1}} \rangle &=& \frac{1}{2}\left(\rho_{y_t=1,x_{t-1}=1} + \rho_{y_t=0,x_{t-1}=0}\right)\\
&=& 1
\end{eqnarray*}
and for $\mathbf{Y}$ causes $\mathbf{X}$ is
\begin{eqnarray*}
\langle \rho_{x_t,y_{t-1}} \rangle &=& \frac{1}{3}\left(\rho_{x_t=1,y_{t-1}=0} \right.\\
& &\left. +\rho_{x_t=0,y_{t-1}=1} + \rho_{x_t=0,y_{t-1}=0}\right)\\
&=& \frac{1}{7}\;\;.
\end{eqnarray*}
The {\em mean observed leaning} that follows from the definition of the mean observed penchants is
\begin{eqnarray}
\label{eqn:meanlean}
\langle \lambda_{y_t,x_{t-1}} \rangle &=& \langle \rho_{y_t,x_{t-1}} \rangle - \langle \rho_{x_t,y_{t-1}} \rangle\\
&=& \frac{6}{7}\;\;.
\end{eqnarray}

The positive leaning implies the probability that $x_{t-1}$ drives $y_t$ is higher than the probability that $y_{t-1}$ drives $x_{t}$; i.e.,\ $\mathbf{X}\xrightarrow{lean}\mathbf{Y}$ given the 1-standard cause-effect assignment.  This result is expected and agrees with the intuitive definition of causality in this example.  

The {\em weighted mean observed penchant} is defined similarly to the mean observed penchant, but each penchant is weighted by the number of times it appears in the data; e.g.,\
\begin{eqnarray*}
\langle \rho_{y_t,x_{t-1}} \rangle_w &=& \frac{1}{L}\left(n_{y_t=1,x_{t-1}=1}\rho_{y_t=1,x_{t-1}=1} \right.\\
& & \left.+ n_{y_t=0,x_{t-1}=0}\rho_{y_t=0,x_{t-1}=0}\right)\\
&=& 1
\end{eqnarray*}
and
\begin{eqnarray*}
\langle \rho_{x_t,y_{t-1}} \rangle_w &=& \frac{1}{L}\left(n_{x_t=1,y_{t-1}=0}\rho_{x_t=1,y_{t-1}=0} \right.\\
& & +n_{x_t=0,y_{t-1}=1}\rho_{x_t=0,y_{t-1}=1}\\
& & \left.+ n_{x_t=0,y_{t-1}=0}\rho_{x_t=0,y_{t-1}=0}\right)\\
&=& \frac{3}{63}\;\;,
\end{eqnarray*}
where $n_{a,b}$ is the number of times the assumed cause $a$ appears with the assumed effect $b$ and $L$ is the library length of $\tilde{\mathbf{X}}$ (i.e., $L=N-l$ where $N$ is the library length of $\mathbf{X}$ and $l$ is the lag used in the $l$-standard cause-effect assignment).  

The {\em weighted mean observed leaning} follows naturally as
\begin{eqnarray*}
\langle \lambda_{y_t,x_{t-1}} \rangle_w &=& \langle \rho_{y_t,x_{t-1}} \rangle_w - \langle \rho_{x_t,y_{t-1}} \rangle_w\\
&=& \frac{60}{63}\;\;.
\end{eqnarray*}
For this example, $\langle \lambda_{y_t,x_{t-1}} \rangle_w\Rightarrow \mathbf{X}\xrightarrow{lean}\mathbf{Y}$ as expected.

Conceptually, the weighted mean observed penchant is preferred to the mean penchant because it accounts for the frequency of observed cause-effect pairs within the data, which is assumed to be a predictor of causal influence.  For example, given some pair $\{\mathbf{A},\mathbf{B}\}$, if it is known that $a_{t-1}$ causes $b_{t}$ and both $b_t = 0\; |\; a_{t-1} = 0$ and $b_t = 0\; |\; a_{t-1} = 1$ are observed, then comparison of the frequencies of occurrence is used to determine which of the two pairs represents the cause-effect relationship.

For this example, the weighted mean observed leaning provides the same causal inference as the mean observed leaning.  The weighted mean calculation will be used in the examples of the following sections.

\subsection{Unobserved penchants}
The {\em unobserved} penchants for $l=1$ for $\mathbf{X}$ causes $\mathbf{Y}$ are
\begin{eqnarray*}
\rho_{y_t=1,x_{t-1}=0} &=& -1\\
\rho_{y_t=0,x_{t-1}=1} &=& -1
\end{eqnarray*}
and for $\mathbf{Y}$ causes $\mathbf{X}$ is
\begin{equation*}
\rho_{x_t=1,y_{t-1}=1}=-\frac{3}{7}\;\;.
\end{equation*}
These values can incorporated into the averaging calculation to yield a {\em mean total penchant}; i.e., for $\mathbf{X}$ causes $\mathbf{Y}$
\begin{eqnarray*}
\langle\langle \rho_{y_t,x_{t-1}} \rangle\rangle &=& \frac{1}{4}\left(\rho_{y_t=1,x_{t-1}=1} + \rho_{y_t=0,x_{t-1}=0}\right.\\
& &\left. \rho_{y_t=1,x_{t-1}=0} + \rho_{y_t=0,x_{t-1}=1}\right) \\
&=& 0
\end{eqnarray*}
and for $\mathbf{Y}$ causes $\mathbf{X}$
\begin{eqnarray*}
\langle\langle \rho_{x_t,y_{t-1}} \rangle\rangle &=& \frac{1}{4}\left(\rho_{x_t=1,y_{t-1}=1} + \rho_{x_t=0,y_{t-1}=0}\right.\\
& &\left. \rho_{x_t=1,y_{t-1}=0} + \rho_{x_t=0,y_{t-1}=1}\right)\\
&=& 0\;\;.
\end{eqnarray*}
Thus, the {\em mean total leaning} (defined analogous to Eqn.\ \ref{eqn:meanlean}) is $\langle\langle \lambda_{y_t,x_{t-1}} \rangle\rangle = \langle\langle \rho_{y_t,x_{t-1}} \rangle\rangle -  \langle\langle \rho_{x_t,y_{t-1}} \rangle\rangle = 0$.  No causal inference can be made with a leaning of zero because it implies $\langle\langle \rho_{y_t,x_{t-1}} \rangle\rangle =  \langle\langle \rho_{x_t,y_{t-1}} \rangle\rangle$.  Thus $\mathbf{X}$ does not have a higher penchant to drive $\mathbf{Y}$ than $\mathbf{Y}$ does to drive $\mathbf{X}$, given the cause-effect assignment used in the leaning calculation.  Such a conclusion would not be useful for causal inference, which implies the mean total leaning is not useful for causal inference in this example.

\subsection{Cause-effect assignment independence}
The causal inference above assumed a cause-effect relationship was known to be correct.  It can be shown, however, that causal inference is independent of the assumed cause-effect relationship.  For example, consider the cause-effect assignment $\{C,E\}=\{y_{t-l},x_t\}$ with $l=1$. The mean observed leaning is
\begin{eqnarray*}
\langle \lambda_{x_t,y_{t-1}} \rangle &=& \langle \rho_{x_t,y_{t-1}} \rangle - \langle \rho_{y_t,x_{t-1}} \rangle\\
&=& -\frac{6}{7}\;\;,
\end{eqnarray*}
which implies $\mathbf{X}\xrightarrow{lean}\mathbf{Y}$, as expected for this example.

In general, $\lambda_{AB} := \rho_{AB} - \rho_{BA}\Rightarrow -\lambda_{AB} = \rho_{BA} - \rho_{AB} := \lambda_{BA}$.  Thus, the causal inference is independent of which times series is initially assumed to be the cause (or effect).  

\subsection{Tolerance domains}
\label{sec:tol}
If the example time series contained noise, then a realization of the example time series $\{\mathbf{X}^\prime,\mathbf{Y}^\prime\}$ could be
\begin{eqnarray*}
\mathbf{X}^\prime &=& \{x_t^\prime\; | \; t=0,1,2,\ldots,9\}\\
&=& \left\{0,0,1.1,0,0,1,-0.1,0,0.9,0\right\}\\
\mathbf{Y}^\prime &=& \{y_t^\prime\; | \; t=0,1,2,\ldots,9\}\\
&=& \left\{0,-0.2,0.1,1.2,0,0.1,0.9,-0.1,0,1\right\}.
\end{eqnarray*}

The previous time series pair, $\{\mathbf{X},\mathbf{Y}\}$ had only five observed penchants, but $\{\mathbf{X}^\prime,\mathbf{Y}^\prime\}$ has more due to the noise.  It can be seen in the time series definitions that $x_t^\prime = x_t \pm 0.1 := x_t \pm \delta_x$ and $x_t^\prime = x_t \pm 0.2 := x_t \pm \delta_y$.  The weighted mean observed leaning for $\{\mathbf{X}^\prime,\mathbf{Y}^\prime\}$ is $\langle \lambda_{y_t^\prime,x_{t-1}^\prime} \rangle_w \approx 0.19$. 

If the noise is not restricted to a small set of discrete values, then the effects of noise on the leaning calculations can be addressed by using the tolerances $\delta_x$ and $\delta_y$ in the probability estimations from the data.  For example, the penchant calculation in Eqn.\ \ref{eqn:rhoex1} relied on estimating $P(y_t=1|x_{t-1}=1)$ from the data, but if, instead, the data is known to be noisy, then the relevant probability estimate may be $P(y_t\in[1-\delta_y,1+\delta_y]|x_{t-1}\in[1-\delta_x,1+\delta_x])$.

If the tolerances, $\delta_x$ and $\delta_y$, are made large enough, then the noisy system weighted mean observed leaning, $\langle \lambda_{y_t^\prime\pm\delta_y,x_{t-1}^\prime\pm\delta_x} \rangle_w$, can, at least in the simple examples considered here, be made equal to the noiseless system weighted mean observed leaning, i.e.,\ $\langle \lambda_{y_t^\prime\pm\delta_y,x_{t-1}^\prime\pm\delta_x} \rangle_w = \langle \lambda_{y_t,x_{t-1}} \rangle_w$.

Tolerance domains, however, can be set too large.  If the tolerance domain is large enough to encompass every point in the time series, then the probability of the assumed cause becomes one, which leads to undefined penchants.  For example, given the symmetric definition of the tolerance domain used in this section, $\delta_x = 2$ implies $P(x_{t-1} = 1\pm\delta_x) = 1$, which implies $\langle \lambda_{y_t^\prime,x_{t-1}} \rangle_w$ is undefined.

This example was used to motivate the need for an understanding of the noise in the measurements, which may not always be possible.  If little is known about the noise, one strategy is to calculate the leanings with several different tolerances, increasing the size of the tolerance domains to the point where the penchants become undefined, and finding the tolerance domains for which the leaning changes sign.  The sizes of these domains can then be compared to suspected noise levels.  This strategy, and others, will be considered in more detail in the following sections.  If the noise level is known, then the task becomes much simpler and the tolerances should just be set to the known (or estimated) noise levels for the individual time series.

As mentioned in Section \ref{sec:lean}, the probabilities required for the leaning calculations are estimated in this example (and all the following ones) through straightforward counting of the data.  As such, the tolerance domains may be thought of as bin widths for the probability estimations.  Work has been done on optimal and data-driven bin width selection procedures (see, e.g., \cite{Wand1997,Scott1979}), usually in the context of finding histograms.  Tolerance domains, however, may be thought of in terms of the causal inference for which the leaning is intended.  The tolerance domain for the ``cause'' (or ``effect'') is the domain in which an analyst considers the data may still reasonably be identified as a ``cause'' (or ``effect'').  It is not required to be symmetric, and the tolerance domain for one time series is not required to be equal to the tolerance domain for the other (which is seen in the example above and most of the examples that follow).

\subsection{Stationarity dependence}
Both $\mathbf{X}$ and $\mathbf{Y}$ are stationary in the original example time series pair $\{\mathbf{X},\mathbf{Y}\}$.  Suppose 1,000 zeros are appended to the end of each of these time series.  The additional zeros in the times series may intuitively seem to make causal inference more difficult.  The probabilities required for the penchant $\rho_{y_t=1|x_{t-1}=1}$ become
\begin{eqnarray*}
P\left( y_t = 1 | x_{t-1} = 1\right) &=& \frac{3}{3} = 1\;\;,\\
P\left( y_t = 1\right) &=& \frac{3}{1009}\;\;,
\end{eqnarray*}
and
\begin{equation*}
P\left( x_{t-1} = 1\right) = \frac{3}{1009}\;\;.
\end{equation*}
These probabilities have become much smaller but the penchant remains the same.  The same is true for $\rho_{y_t=0|x_{t-1}=0}$.  Despite the additional zeros, $\mathbf{Y}$ can still only take the values 1 or 0.  This knowledge along with the above penchants implies $n_{y_t=1,x_{t-1}=1}+n_{y_t=0,x_{t-1}=0}=L$, which implies $\langle \rho_{y_t,x_{t-1}} \rangle_w = 1$.  The other three observed penchants, however, do change as a result of the appended zeros.  Previously, $|\rho_{x_t=1,y_{t-1}=0}| = |\rho_{x_t=0,y_{t-1}=1}| = |\rho_{x_t=0,y_{t-1}=0}| = 3/7$, but with the appended zeros, $|\rho_{x_t=1,y_{t-1}=0}| = |\rho_{x_t=0,y_{t-1}=1}| = |\rho_{x_t=0,y_{t-1}=0}| = 3/1009$.  The weighted mean observed leaning, $\langle \lambda_{y_t,x_{t-1}} \rangle_w$, changes from $60/63$ to approximately $1012/1009$ because of the appended zeros.  This value is higher than the previous value but yields the same causal inference.  

Consider another non-stationary times series pair, $\{\mathbf{X}_L,\mathbf{R}_L\}$, where the non-stationary response signal is $\mathbf{R}_L = \{0,0,0,1,1,1,2,2,2,3\}$.  The weighted mean observed leaning calculated under the 1-standard assignment with no tolerance domains still leads to a causal inference that agrees with intuition; i.e.\ $\langle \lambda_{r_t,x_{t-1}} \rangle_w \approx 0.11 \Rightarrow \mathbf{X}_L\xrightarrow{lean}\mathbf{R}_L$ as expected.  This result, however, depends on the library length of the data.

$\{\mathbf{X}_L,\mathbf{R}_L\}$ is a specific instance of the following time series pair:
\begin{eqnarray}
\left\{\mathbf{X},\mathbf{R}\right\} = \left\{\{x_t\},\{r_t\}\right\}
\end{eqnarray}
where $t=0,1,2,\ldots,L$,
\begin{equation}
x_t = \left\{
  \begin{array}{lr}
    0 & \forall\; t\in\{t\;|\;t\bmod 3 \neq 0\}\\
    1 & \forall\; t\in\{t\;|\;t\bmod 3 = 0\}
  \end{array}
\right.
\end{equation}
and
\begin{equation}
r_t = x_{t-1}+r_{t-1}
\end{equation}
with $r_0 = 0$.  The weighted mean observed leaning, under the 1-standard assignment with no tolerance domains, for $\{\mathbf{X},\mathbf{R}\}$ depends on $L$.  As $L$ is increased, the leaning calculation will eventually lead to causal inferences that do not agree with intuition; e.g.,\ $L = 20 \Rightarrow \langle \lambda_{r_t,x_{t-1}} \rangle_w \approx 1.8\times10^{-3} \Rightarrow \mathbf{X}\xrightarrow{lean}\mathbf{R}$ and $L = 50 \Rightarrow \langle \lambda_{r_t,x_{t-1}} \rangle_w \approx -2.5\times10^{-3} \Rightarrow \mathbf{R}\xrightarrow{lean}\mathbf{X}$.  

As $L$ is increased, the number of possible observed effects for a given observed cause increases.  Thus, under the 1-standard assignment $\{C,E\} = \{x_{t-1},r_t\}$, $x_{t-1}=1$ precedes three different values, $r_t = 1$, $2$, and $3$, if $L=10$, but it precedes fifteen different values if $L=50$.  The leaning calculations are methods for counting (in a specific way) the number of times (and ways in which) an observed cause-effect pair appears in the data.  The causal inference becomes more difficult for non-stationary time series pairs because repeated cause-effect pairs in the data may be more rare than in the stationary examples.  This effect is very similar to the effect seen when the impulse signal was noiseless but the response was noisy.  Unfortunately, it cannot be remedied with tolerance domains for the non-stationary case.  For example, for $\{\mathbf{X},\mathbf{R}\}$, the cardinality of the set $\{r_t\;|\;x_{t-1}=1\}\rightarrow\infty$ as $L\rightarrow\infty$, and penchants would not be defined given a tolerance domain for $\mathbf{R}$ of $\delta_r=\infty$.

These shortcomings of the weighted mean observed leaning when applied to non-stationary data, however, do not imply that causal inference of non-stationary data cannot be done using a different application of the observed penchants.  For example, replacing the weighted mean calculation in the weighted mean observed leaning calculation with a median calculation leads to a {\em median observed leaning}, $[\lambda_{r_t,x_{t-1}}] \approx 5.3\times 10^{-3}\Rightarrow \mathbf{X}\xrightarrow{lean}\mathbf{R}$ for $L=50$ as expected, where $[\cdot]$ is used to denote the median.  Of course, even though the median leaning calculation agrees with intuition for a library length where the mean leaning calculation did not, there is no reason to believe the median leaning calculation will not also eventually provide counterintuitive causal inferences as $L$ is increased.  

A more basic strategy for dealing with non-stationary data would be to define the observed penchant using a different cause-effect assignment.  For example, the $l$-standard assignment (with $l=1$) used above, i.e.,\ $\{C,E\}=\{x_{t-1},r_t\}$, might be replaced with an $l$-AR (autoregressive) assignment with $l=1$ of $\{C,E\}=\{(x_{t-1},r_{t-1}),r_t\}$.  An observed penchant may be calculated with an assumed cause of $(x_{t-1}=1,r_{t-1}=0)$ and an assumed effect of $r_t = 1$.  The algorithms to compute the observed penchants from the data become more complicated as the cause-effect assignment becomes more complicated, but the basic definition of the penchant provides a very general conceptual framework for causal inference.

\section{Simple Example Systems}
In this section the weighted mean observed leaning using the $l$-standard cause-effect assignment for various $l$, will be applied to dynamical systems and empirical data sets with known causal relationships.  The usefulness of the leaning as a tool for causal inference is tested directly with synthetic and empirical time series data sets for which there is an intuitive understanding of the driving relationships within the system.

\subsection{Impulse with Noisy Response Linear Example}
\label{sec:IR}
Consider the linear example dynamical system of
\begin{eqnarray}
\label{eqn:IReqn}
\left\{\mathbf{X},\mathbf{Y}\right\} = \left\{\{x_t\},\{y_t\}\right\}
\end{eqnarray}
where $t=0,1,2,\ldots,L$,
\begin{equation*}
x_t = \left\{
  \begin{array}{lr}
    2 & t = 1\\
    0 & \forall\; t\in\{t\;|\;t\neq 1 \;\mathrm{and}\; t\bmod 5 \neq 0\}\\
    2 & \forall\; t\in\{t\;|\;t\bmod 5 = 0\}
  \end{array}
\right.
\end{equation*}
and
\begin{equation*}
y_t = x_{t-1} + B\eta_t
\end{equation*}
with $y_0 = 0$, $B\in\mathbb{R}\ge 0$ and $\eta_t\sim\mathcal{N}\left(0,1\right)$.  Specifically, consider $B\in[0,1]$.  The driving system $\mathbf{X}$ is a periodic impulse with a signal amplitude above the maximum noise level of the response system, and the response system $\mathbf{Y}$ is a lagged version of the driving signal with $\mathcal{N}\left(0,1\right)$ of amplitude $B$ applied at each time step.  
\begin{figure}[ht]
\begin{tabular}{cc}
\includegraphics[scale=0.38]{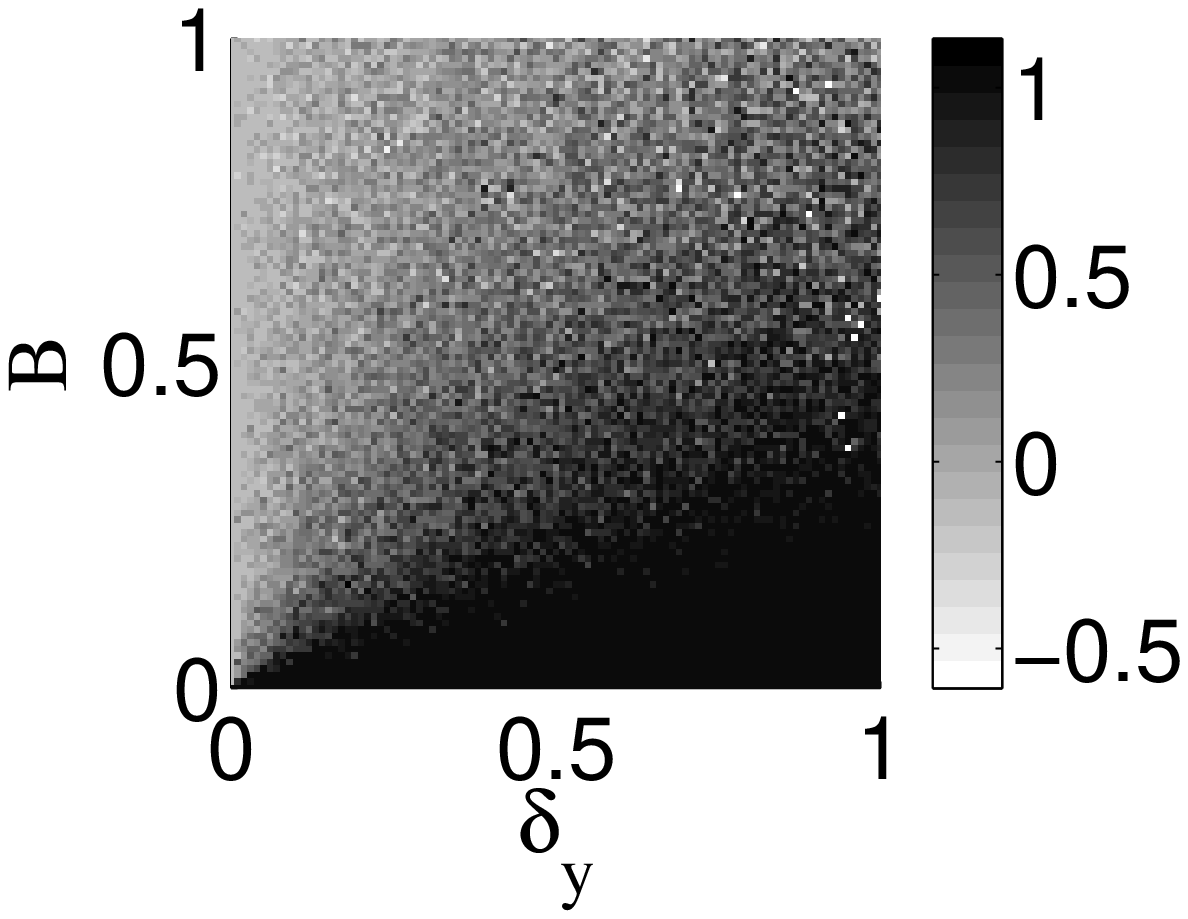} &
\includegraphics[scale=0.38]{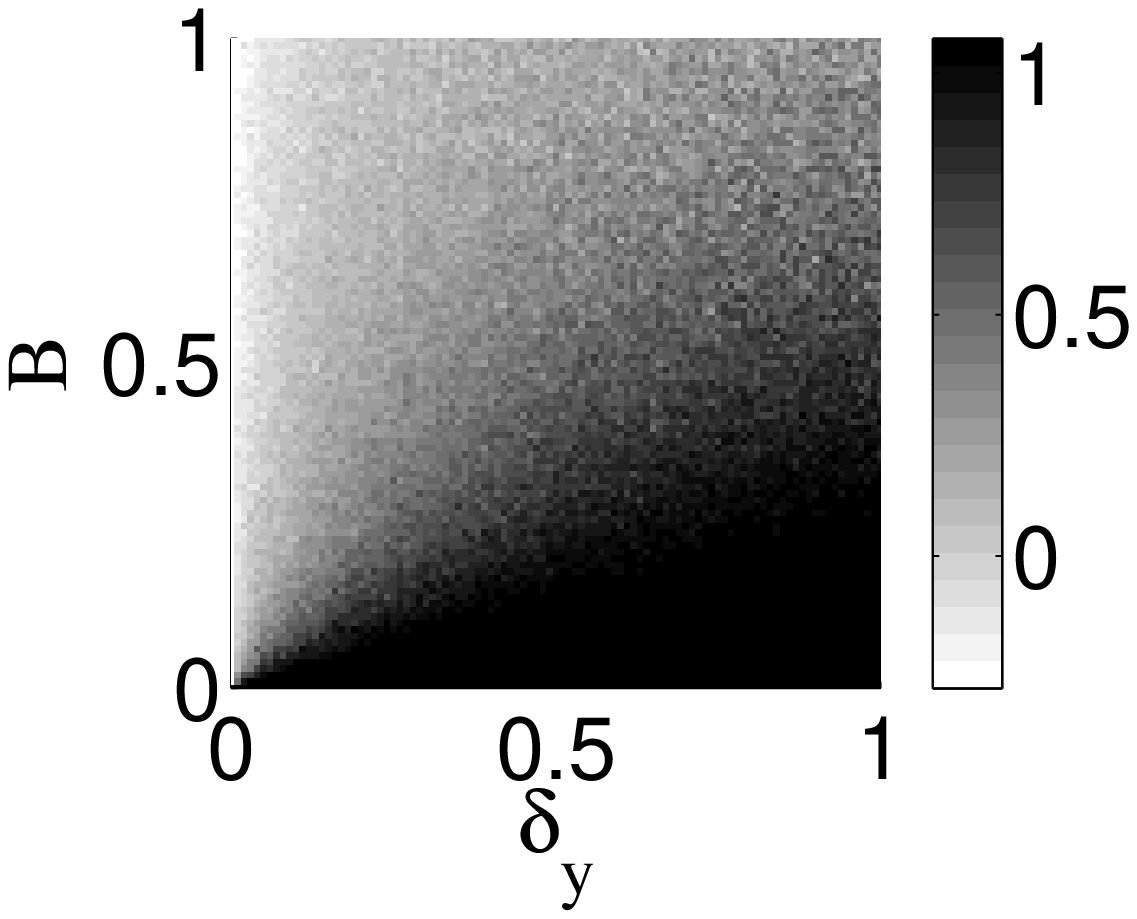} \\
(a) & (b)\\
\includegraphics[scale=0.38]{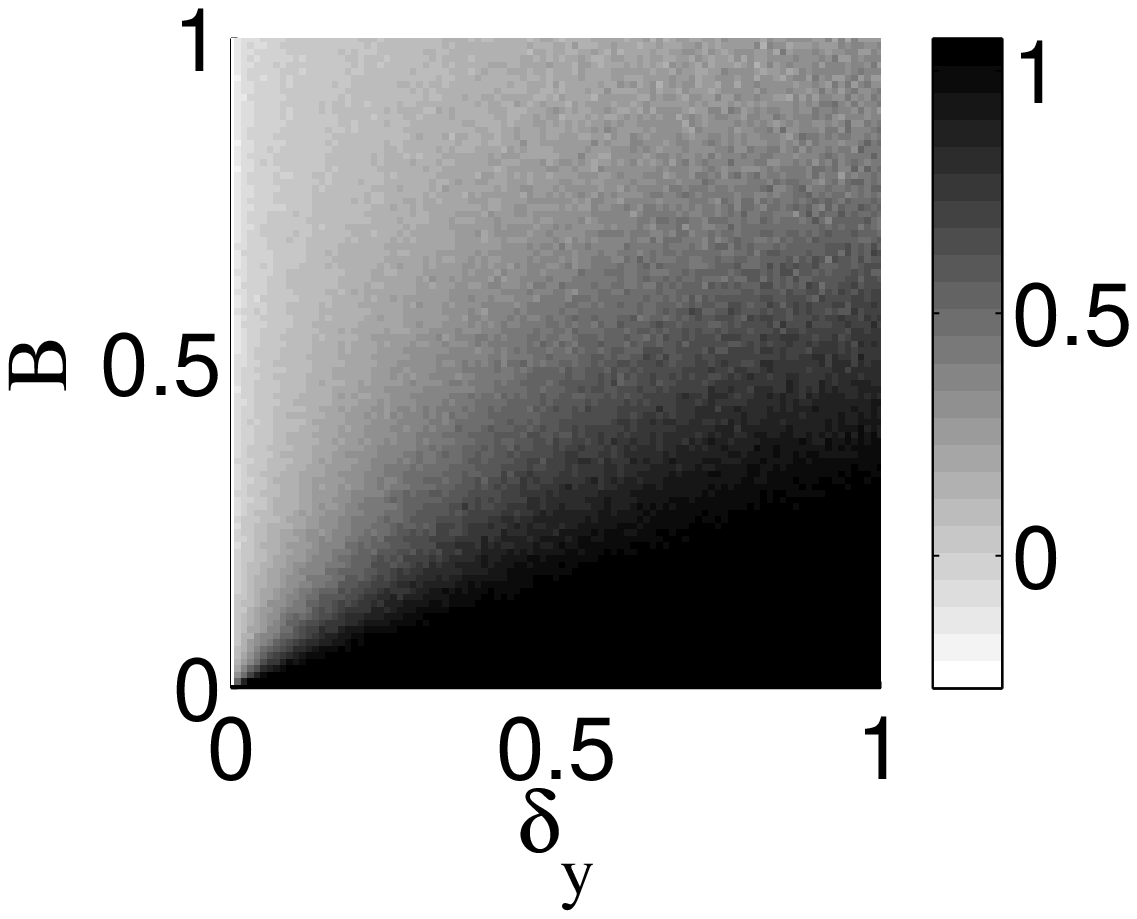} &
\includegraphics[scale=0.38]{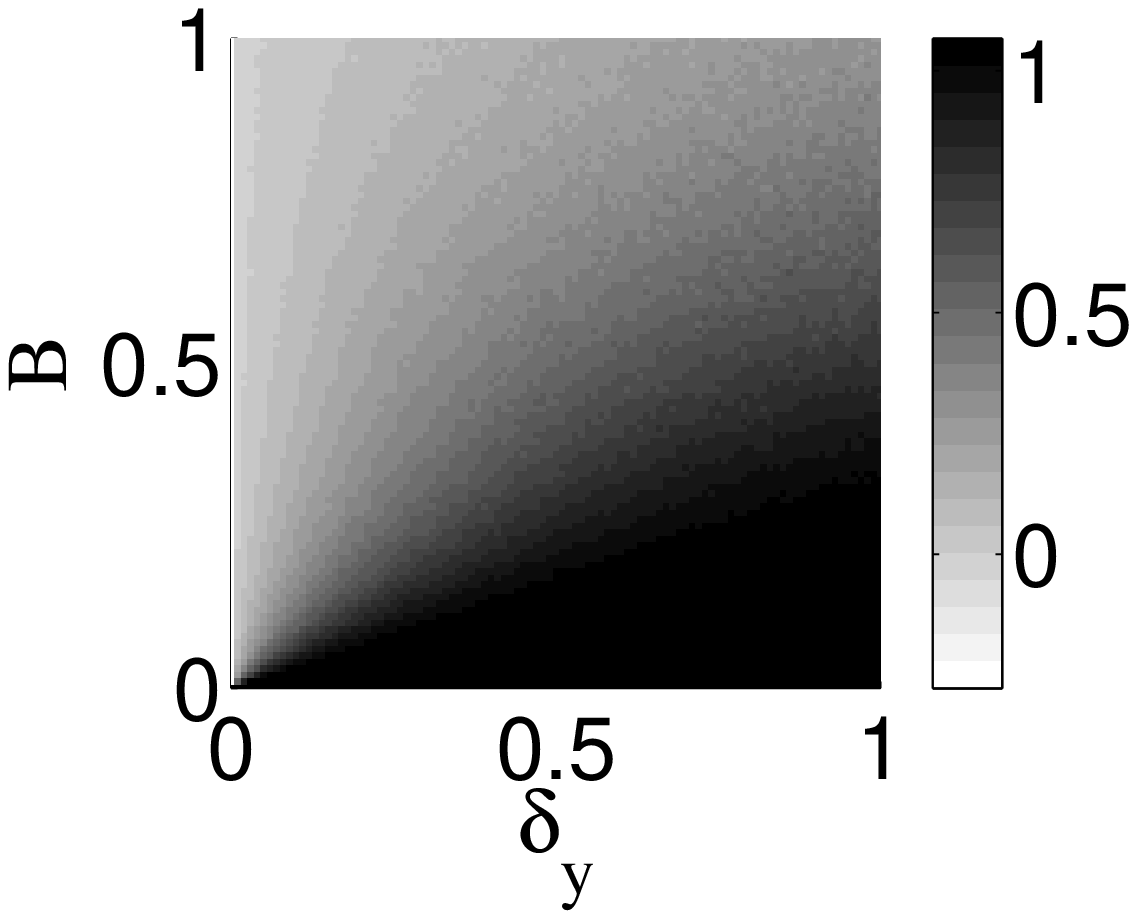} \\
(c) & (d) \\
\end{tabular}
\caption{The unitless leaning of Eqn.\ \ref{eqn:IReqn} is a function of the noise, the tolerance used for terms from $\mathbf{Y}$, and the length of the signals and both $\delta_y$ and $B$ are unitless.  See the text for an explanation of the missing data for large $\delta_y$.  The subplots are each for different signal lengths, (a) $L=10$, (b) $L=50$, (c) $L=250$, and (d) $L=1750$.}
\label{fig:IRexChangeL}
\end{figure}

Figure \ref{fig:IRexChangeL} shows how the weighted mean observed leaning using the 1-standard cause-effect assignment, $\tilde{\lambda}$, changes as the noise amplitude $B$ and tolerance $\delta_y$ are increased in increments of 0.01.  The synthetic data sets $\mathbf{X}$ and $\mathbf{Y}$ are constructed such that intuitively $\mathbf{X}$ drives $\mathbf{Y}$.  Thus, it is expected that $\mathbf{X}\xrightarrow{lean}\mathbf{Y}$ which implies $\tilde{\lambda} > 0$.  Figure \ref{fig:IRexChangeL} shows that this expectation is met except when $\delta_y > B$ even for a short library length of $L=10$.  Examples of undefined penchants due to large tolerance domains, as discussed in section \ref{sec:tol}, are seen as $\delta_y$ is increased in the $L=10$ example.

Figure \ref{fig:IRexChangeL} shows using the strategy of $\delta_y = B$ always leads to causal inferences that agree with intuition for $L>10$ in this example.  However, as discussed in section \ref{sec:tol}, knowing $B$ {\em a priori} may be unrealistic with empirical data sets.  Consider the following three methods for estimating $\delta_y$ from the data:
\begin{enumerate}
\item {\em lagged linear response deviation} - $\delta_y$ is set to the mean absolute deviation of $y_t$ from $x_{t-1}$; i.e.,\ $\delta_y = \langle|y_t-x_{t-1}|\rangle$.
\item {\em normalized standard deviation} - $\delta_y$ is set to the standard deviation of $\{|\mathbf{Y}-\langle\mathbf{Y}\rangle|\}$ where $\langle\mathbf{Y}\rangle$ is the mean of $\mathbf{Y}$; i.e.,\ $\delta_y = \sigma_{|y_t-\langle y_t\rangle|}$.
\item {\em n-bin mean standard deviation} - $\delta_y$ is set to the mean standard deviation of $n$ bins of $\mathbf{Y}$; i.e.,\ $\delta_y = \langle \sigma_{B_i}\rangle$ where $B_i$ is the $i$th bin of an $n$-bin histogram of $\mathbf{Y}$.    
\end{enumerate}  
Table \ref{tab:IRlagTolComp} shows $\tilde{\lambda}$ for instances of Eqn.\ \ref{eqn:IReqn} with $B = 0$, $0.1$, $0.5$, and $0.8$ and $L=100$ (and $n=5$ in method 3).
\begin{table}
\begin{tabular}{l|ccc}
 \multirow{2}{*}{ $B$ }& \multicolumn{3}{|c}{$\tilde{\lambda}$}\\
 \cline{2-4}
 & method 1 & method 2 & method 3\\
\hline
0.0 & 1.0 & 1.0 & 1.0\\
0.1 & 0.40 & 1.0 & 0.48\\
0.5 & 0.39 & 0.79 & 0.26\\
0.8 & 0.30 & 0.44 & 0.10\\
\end{tabular}
\caption{$\tilde{\lambda}$ using three different estimation methods for $\delta_y$: (1) lagged linear response deviation, (2) normalized standard deviation, and (3) n-bin mean standard deviation.}
\label{tab:IRlagTolComp}
\end{table}

The three different methods yield different values for the leaning, but all the methods lead to the same causal inference, $\mathbf{X}\xrightarrow{lean}\mathbf{Y}$, which agrees with intuition.  These methods are meant to be examples of using the data to set $\delta_y$ if $B$ is not known.  These methods are not expected to be reasonable estimates for $\delta_x$ and $\delta_y$ in general.  For example, method 1 assumes a linear relationship between $\mathbf{X}$ and $\mathbf{Y}$ that may be unreasonable to assume in general.  However, Table \ref{tab:IRlagTolComp} shows different methods for setting $\delta_y$ can lead to the same causal inference.  Setting the tolerances requires an understanding of the noise in the times series data.  The leaning is meant to be part of an exploratory causal analysis of the time series data and cannot exist independently of other exploratory analysis of the data, including analysis of the noise levels.  

\subsection{Cyclic Linear Example}
The calculations in the previous subsection were only for the 1-standard assignment ($l=1$) and are expected to be useful for causal inference given Eqn.\ \ref{eqn:IReqn}.  However, deciding which $l$-standard assignment to use given empirical, rather than synthetic, data sets may be more difficult.  It is expected that several different $l$-standard assignments would be used as part of any exploratory causal analysis using leaning.  This section contains an example that plots the leanings for a set of different $l$-assignments and shows the maximum leaning in the set is near the expected value, i.e.,\ near the lag value that appears explicitly in the dynamical system used to create the synthetic data sets. 

Consider the linear example dynamical system of
\begin{eqnarray}
\label{eqn:cyceqn}
\left\{\mathbf{X},\mathbf{Y}\right\} = \left\{\{x_t\},\{y_t\}\right\}
\end{eqnarray}
where $t=0,1,2,\ldots,L$,
\begin{equation*}
x_t = \sin(t)
\end{equation*}
and
\begin{equation*}
y_t = x_{t-1} + B\eta_t
\end{equation*}
with $y_0 = 0$, $B\in[0,1]$ in steps of 0.01 and $\eta_t\sim\mathcal{N}\left(0,1\right)$.  This example is very similar to the previous one, except that the driving system $\mathbf{X}$ is sinusoidal.

\begin{figure}[ht]
\includegraphics[scale=0.47]{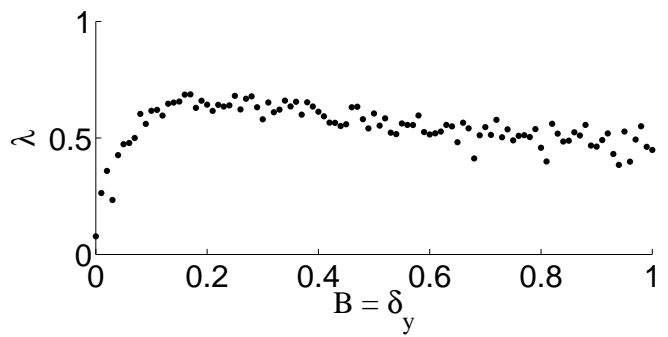}
\caption{Weighted mean observed leaning for Eqn.\ \ref{eqn:cyceqn} and a tolerance for the leaning calculation set to $\delta_y=B$.  $\lambda$ is always positive, which implies $\mathbf{X}\xrightarrow{lean}\mathbf{Y}$.}
\label{fig:cyc1}
\end{figure}

\begin{figure}[ht]
\includegraphics[scale=0.47]{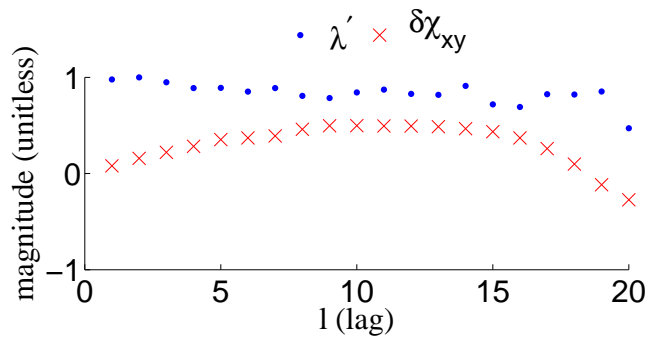}
\caption{(Color available online.) The unitless, normalized leaning, $\lambda^\prime$, of Eqn.\ \ref{eqn:cyceqn} can be plotted for different $l$-standard cause-effect assignments along with the cross correlation, $\chi$ for the same lags, $l$, to show how the two values compare for this simple cyclic example.}
\label{fig:cyc2}
\end{figure}

Figures \ref{fig:cyc1} and \ref{fig:cyc2} were calculated for an instance of Eqn.\ \ref{eqn:cyceqn} with $L=41$ generated by sampling one period of $\mathbf{X}$ with\ $t\in\{0,f\pi,2f\pi,3f\pi,\ldots,2\pi\}$ and $f=1/20$.  Figure \ref{fig:cyc1} shows the weighted mean observed leaning using the 1-standard assignment, $\lambda$, is always positive given $\delta_y=B$.  So, as was seen in the previous example, the leaning implies $\mathbf{X}\xrightarrow{lean}\mathbf{Y}$, which agrees with intuition for this example. 

The driving relationship in this example can be difficult to discern using unmodified CCM techniques \cite{Weigel2014}.  It has been argued that lagged cross-correlation techniques are the preferred causal inference tool in most situations because of their simplicity \cite{McNames2007}.  The lagged cross-correlation is defined as
\begin{equation}
\chi_{xy}^l = \frac{\mathrm{E}\left[\left(x_t-\mu_x\right)\left(y_{t-l}-\mu_y\right)\right]}{\sigma_x\sigma_y}\;\;,
\end{equation}
where $\mathrm{E}[z_t]$ is the expectation value of $\{z_t\}$, $\mu_{x(y)}$ is the mean of $\mathbf{X}$ ($\mathbf{Y}$), and $\sigma_{x(y)}$ is the standard deviation of $\mathbf{X}$ ($\mathbf{Y}$).  The cross-correlation is often used for causal inference by introducing a difference quantity \cite{Rogosa1980}
\begin{equation}
\delta\chi_{xy}^l = \chi_{xy}^l - \chi_{yx}^l\;\;.
\end{equation}
The sign of $\delta\chi_{xy}^l$ is used, similar to the leaning approach, to determine the causal inference; i.e.,\  $\delta\chi_{xy}^l>0$ implies $\mathbf{X}$ ``causes'' $\mathbf{Y}$ and $\delta\chi_{xy}^l<0$ implies $\mathbf{Y}$ ``causes'' $\mathbf{X}$ \cite{Rogosa1980}.  

Figure \ref{fig:cyc2} shows how $\delta\chi_{xy}^l$ compares to the leaning given $l\in[1,21]$ for an instance of Eqn.\ \ref{eqn:cyceqn} with $B=0.5$.  In Figure \ref{fig:cyc2}, the leaning has been normalized for presentation clarity as
\begin{equation}
\lambda^\prime = \frac{\lambda_l}{\max_{l\in[1,21]} \lambda_l}\;\;,
\end{equation}
where $\lambda_l$ is the weighted mean observed leaning using the $l$-standard assignment ($\lambda_1$ is plotted in Figure \ref{fig:cyc1}).  The maximum leaning given $l\in[1,20]$ is approximately 0.625, so the normalized leanings shown in Figure \ref{fig:cyc2} have a scaling factor of approximately 1.6.  

Both $\lambda^\prime$ and $\delta\chi_{xy}^l$ lead to the same causal inference, i.e.\ $\mathbf{X}$ ``drives'' $\mathbf{Y}$, for $l\in[1,19]$, although only the leaning agrees with intuition for $l=20$ and $l=21$ in this example.  Thus, both tools agree with intuition for small lags in this simple cyclic example.  The leaning, however, has its maximum values near the smallest lags, which is expected given Eqn.\ \ref{eqn:cyceqn}, while the cross-correlation difference has its maximum values at lags that do not explicitly appear in Eqn.\ \ref{eqn:cyceqn}.  

\subsection{RL Circuit Example}
\label{sec:rlcirc}

The cross-correlation difference technique is also known to be unreliable given nonlinear dynamics \cite{Rogosa1980}.  Leanings of data sets generated from nonlinear dynamics will be discussed in Section \ref{sec:nonli}.  Neither of the previous examples has been physically motivated, so this section discusses exploratory causal inference of synthetic data sets generated from the well-known dynamics of a physical system.

Consider a series circuit containing a resistor, inductor, and time varying voltage source related by
\begin{equation}
\label{eqn:it}
\frac{dI}{dt} = \frac{V(t)}{L} - \frac{R}{L} I(t),
\end{equation}
where $I(t)$ is the current at time $t$, $V(t)= \sin\left(t\right)$ is the voltage at time $t$, $R$ is the resistance, and $L$ is the inductance.  The time series pair for this example is then 
\begin{eqnarray}
\label{eqn:RLcirceqn}
\left\{\mathbf{V},\mathbf{I}\right\} = \left\{\{V_t\},\{I_t\}\right\}
\end{eqnarray}
where $\mathbf{V}$ is the set of discrete values of $V(t)$ evaluated using $t\in\{0,f\pi,2f\pi,3f\pi,\ldots,8\pi\}$ with $f=1/10$ and $\mathbf{I}$ is the set of discrete values found either by solving Eqn.\ \ref{eqn:it} numerically or by evaluating the analytical solution 
\begin{equation}
I(t) = \frac{L}{D}e^{-\frac{t}{\tau}}+\frac{R}{D}\sin(t)-\frac{L}{D}\cos(t)
\end{equation}
with $D = L^2 + R^2$ and $\tau = L/R$, for the same time set used for $\mathbf{V}$.

Physical intuition is that $V$ drives $I$, and so we expect to find that $\mathbf{V}\xrightarrow{lean}\mathbf{I}$.  The weighted mean observed leaning using the 1-standard assignment, $\lambda_1$, can be used to test this expectation.  Unlike the previous examples, however, there is no noise term in the dynamics (such as $B$ in Eqn.\ \ref{eqn:IReqn} and \ref{eqn:cyceqn}), so setting the tolerance domains, e.g., $\delta_I$, will not be as straightforward.  

Table \ref{tab:rl} shows $\lambda_1$ for both the analytical solution and a numerical solution to Eqn. \ref{eqn:it} using the {\em ode45} integration function in MATLAB.  The time series $\mathbf{V}$ is created by defining values at fixed points and using linear interpolation to find the time steps required by the ODE solver.  Two different physical scenarios are considered in which $L$ and $R$ are constant, $L=10$ H and $R=5$ $\Omega$ and $L=5$ H and $R=20$ $\Omega$.  
\begin{table*}
\begin{tabular}{cc}
\begin{tabular}{lcc}
$\delta_I$ & $\lambda_1$ ({\em ode45}) & $\lambda_1$ (analytical)\\
\hline
0 & -0.132 & -0.089 \\
$10^{-6}$ & -0.132 & 0.493 \\
$10^{-5}$ & -0.108 & 0.548 \\
$10^{-4}$ & 0.188 & 0.564 \\
$10^{-3}$ & 0.582 & 0.581 \\ 
$10^{-2}$ & 0.730 & 0.727 \\
$10^{-1}$ & {\em undefined} & {\em undefined}\\
\end{tabular} &
\begin{tabular}{lcc}
$\delta_I$ & $\lambda_1$ ({\em ode45}) & $\lambda_1$ (analytical)\\
\hline
0 & -0.132 & -0.132 \\
$10^{-6}$ & -0.132 & -0.132 \\
$10^{-5}$ & -0.120 & -0.096 \\
$10^{-4}$ & 0.011 & 0.098 \\
$10^{-3}$ & 0.398 & 0.386 \\
$10^{-2}$ & 0.676 & 0.675 \\
$10^{-1}$ & 0.314 & 0.315 \\
\end{tabular} \\
(a) $R = 20$ $\Omega$, $L = 5$ H & (b) $R = 5$ $\Omega$, $L = 10$ H\\
\end{tabular}
\caption{The leaning $\lambda_1$ depends on both $\delta_I$ and the method for computing $\mathbf{I}$ in this example.  These two cases show that the values of $\delta_I$ for which the leaning starts to agree with intuition can also depend on the physical system parameters (e.g., $\tau$).}
\label{tab:rl}
\end{table*}

The previously discussed strategy of increasing $\delta_I$ until the leaning becomes undefined and then reporting the leaning calculated using the largest $\delta_I$ for which it is defined would lead to a causal inference that agrees with intuition for this example.  Specifically, from Table \ref{tab:rl}(a) $\delta_I=10^{-2}\Rightarrow\lambda_1\approx 0.7\Rightarrow\mathbf{V}\xrightarrow{lean}\mathbf{I}$, as expected.  

Discussion on setting the tolerance domains has centered on understanding the noise in the system.  This example illustrates that the ``noise'' being considered does not need to be a physical noise source in the system (there are no explicit noise terms in Eqn.\ \ref{eqn:RLcirceqn}).  For example, the numerical tolerance of the ODE solver was set to $10^{-3}$ for the results shown in both Table \ref{tab:rl}, and for both examples setting $\delta_I=10^{-3}$ would lead to causal inferences that agree with intuition.  

Consider, for example, the peak values of $\mathbf{V}$.  The time steps of these peaks are $\mathbf{T}_{peak} = \{t|V_t =1\} = \{6,26,46,66\}$.  The values of $\mathbf{I}$ given $\tau = 0.25$ that immediately follow these peaks are $\mathbf{I}^{peak}_{0.25} = \{I_t|t\in\{7,27,47,67\}\}$.  The peak values associated to $\tau=2$ can also be found.  The standard deviation of the first set is $\sigma_{0.25}^{peak} \approx 10^{-6}$ and the standard deviation of the second set is $\sigma_{2}^{peak} \approx 10^{-2}$.  Table \ref{tab:rl}(a) (for $\sigma_{0.25}^{peak}$) and (b) (for $\sigma_{2}^{peak}$) shows setting $\delta_I$ to the appropriate standard deviation of the peaks would lead to causal inferences that agree with intuition.  Rather than physical noise levels, the noise levels used to set the tolerance domains for the leaning calculations is better thought of as the spread in the possible values of an assumed effect that may reasonably be considered due to the same assumed cause.  
    
This example can also illustrate the importance of sample frequency and sample length.  The leaning calculation requires an assumed cause and effect pair to appear in the data enough times to provide a reliable estimates of probabilities.  Thus, data that is sampled for too few periods or too sparsely can lead to counter-intuitive leanings.  For example, if there is only a single peak in the assumed driving time series because of poor sampling, then there can only be a single response value, which would be insufficient to reliably provide the conditional probabilities in the leaning calculation for that assumed cause-effect pair.  For Eqn.\ \ref{eqn:RLcirceqn} with the analytical solution for $\mathbf{I}$, if $\delta_I=10^{-3}$ and $\tau = 0.25$, then $t\in\{0,f\pi,2f\pi,3f\pi,\ldots,2\pi\}$ with $f=1/10$ leads to $\lambda_1 = -0.045$ and $t\in\{0,f\pi,2f\pi,3f\pi,\ldots,3\pi\}$ with $f=2/3$ leads to $\lambda_1 = -0.167$, both of which disagree with intuition.

\subsection{Nonlinear Example}
\label{sec:nonli}

The examples so far have all had a linear relationship between the driving signal and the response signal.  Of the four broad categories of time series causality tools, transfer entropy \cite{kaiser2002} and SSR methods \cite{Sugihara2012} are the two categories that can be applied to nonlinear data sets without modification.  The conceptual framework of Granger causality is not restricted by the linearity of the data set \cite{Granger1980}, but the original formulation by Granger must be modified to do so \cite{Sun2008}.  Lagged cross-correlation techniques are known to be unreliable if the data sets are generated by nonlinear dynamics \cite{Rogosa1980}.  The next examples are for nonlinear systems.

Consider the nonlinear dynamical system of
\begin{eqnarray}
\label{eqn:nonlinearEX}
\left\{\mathbf{X},\mathbf{Y}\right\} = \left\{\{x_t\},\{y_t\}\right\}
\end{eqnarray}
where $t=0,1,2,\ldots,L$,
\begin{equation*}
x_t = \sin(t)
\end{equation*}
and
\begin{equation*}
y_t = Ax_{t-1}\left(1-Bx_{t-1}\right)+C\eta_t,
\end{equation*}
with $y_0 = 0$, with $A,B,C\in[0,1]$ and $\eta_t\sim\mathcal{N}\left(0,1\right)$ with $t\in\{0,f\pi,2f\pi,3f\pi,\ldots,6\pi\}$ and $f=1/30$ so that $L=181$.

\begin{figure}[ht]
\begin{tabular}{cc}
\includegraphics[scale=0.34]{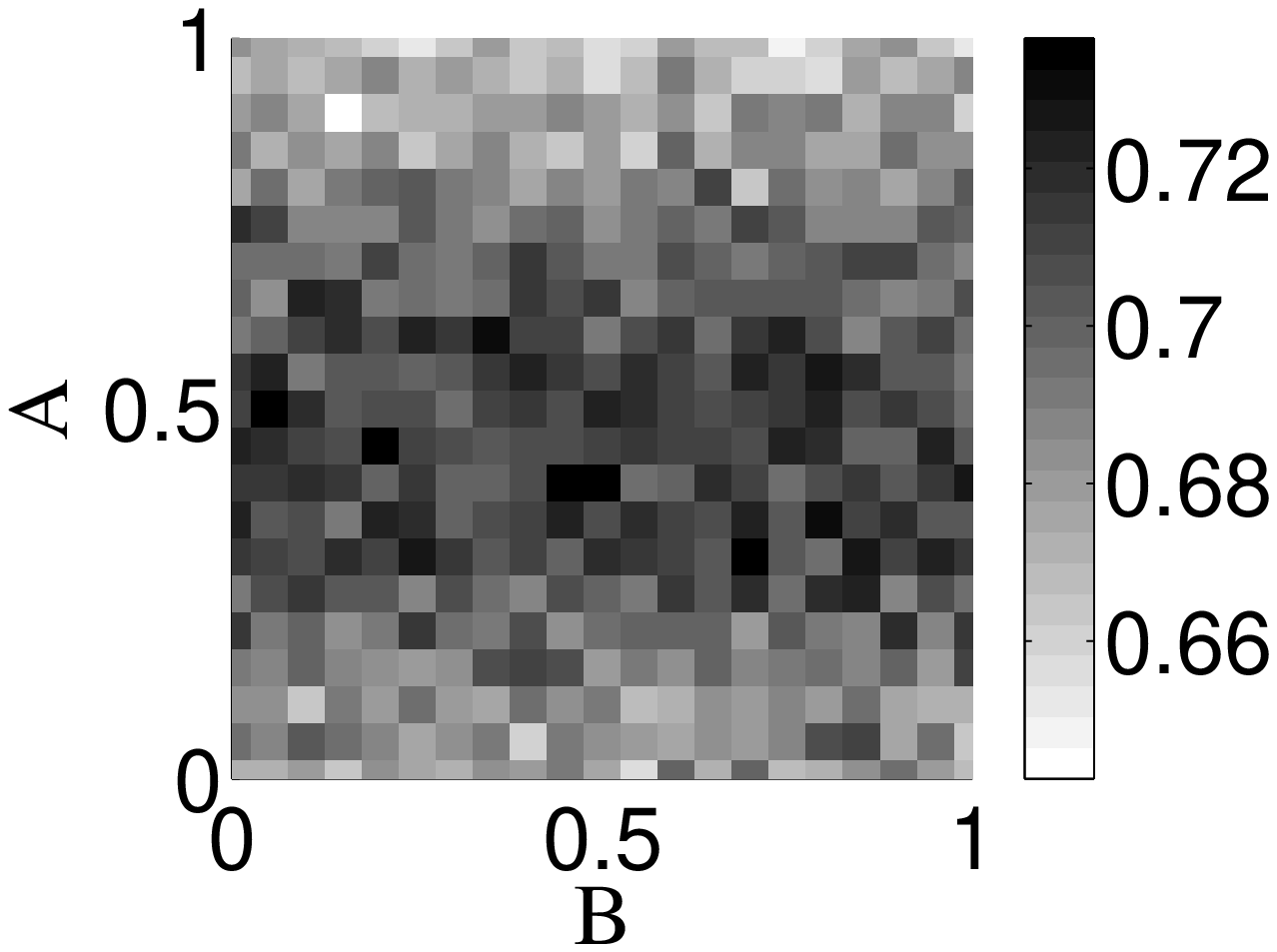} &
\includegraphics[scale=0.34]{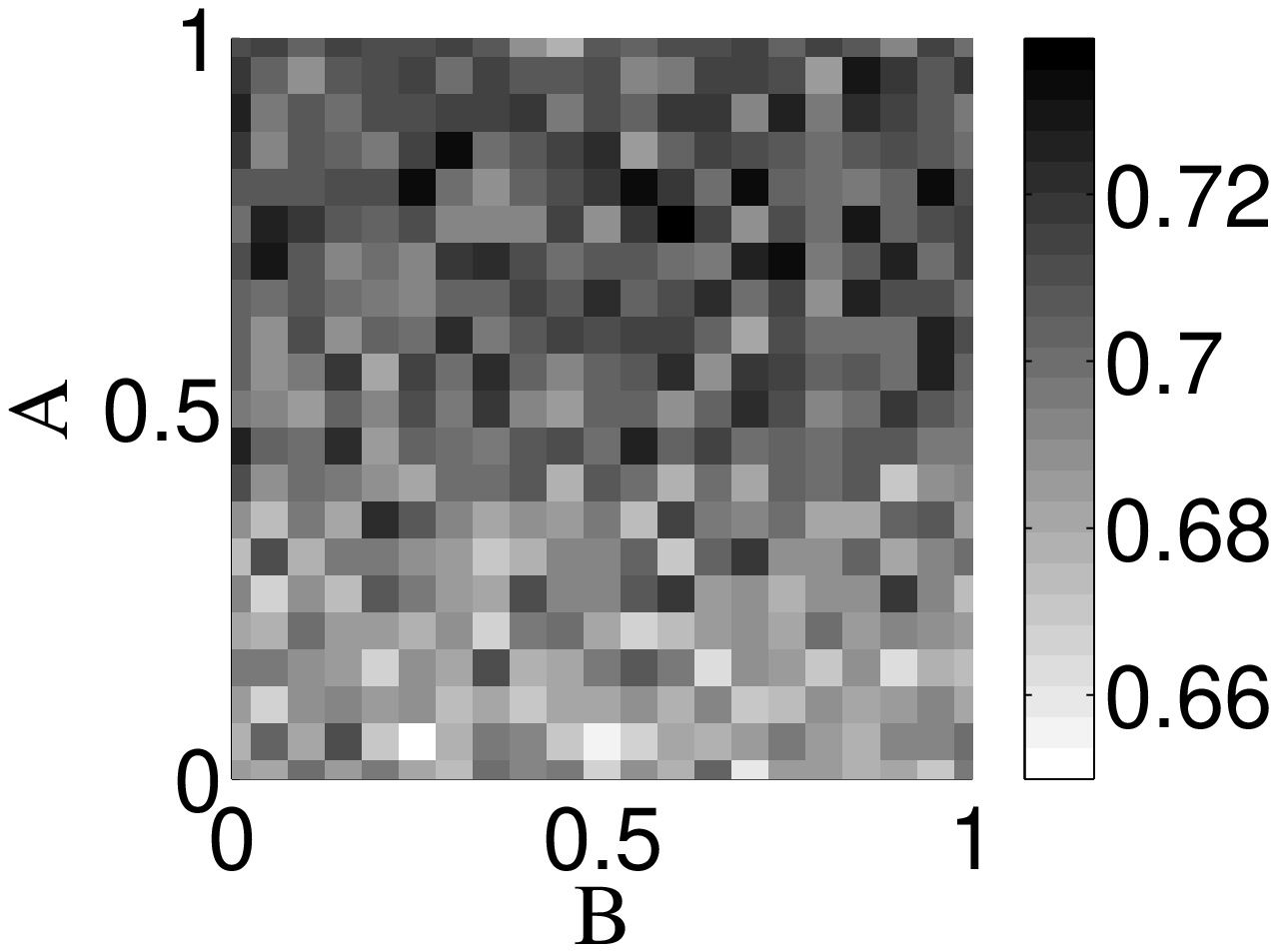} \\
(a) & (b)\\
\includegraphics[scale=0.34]{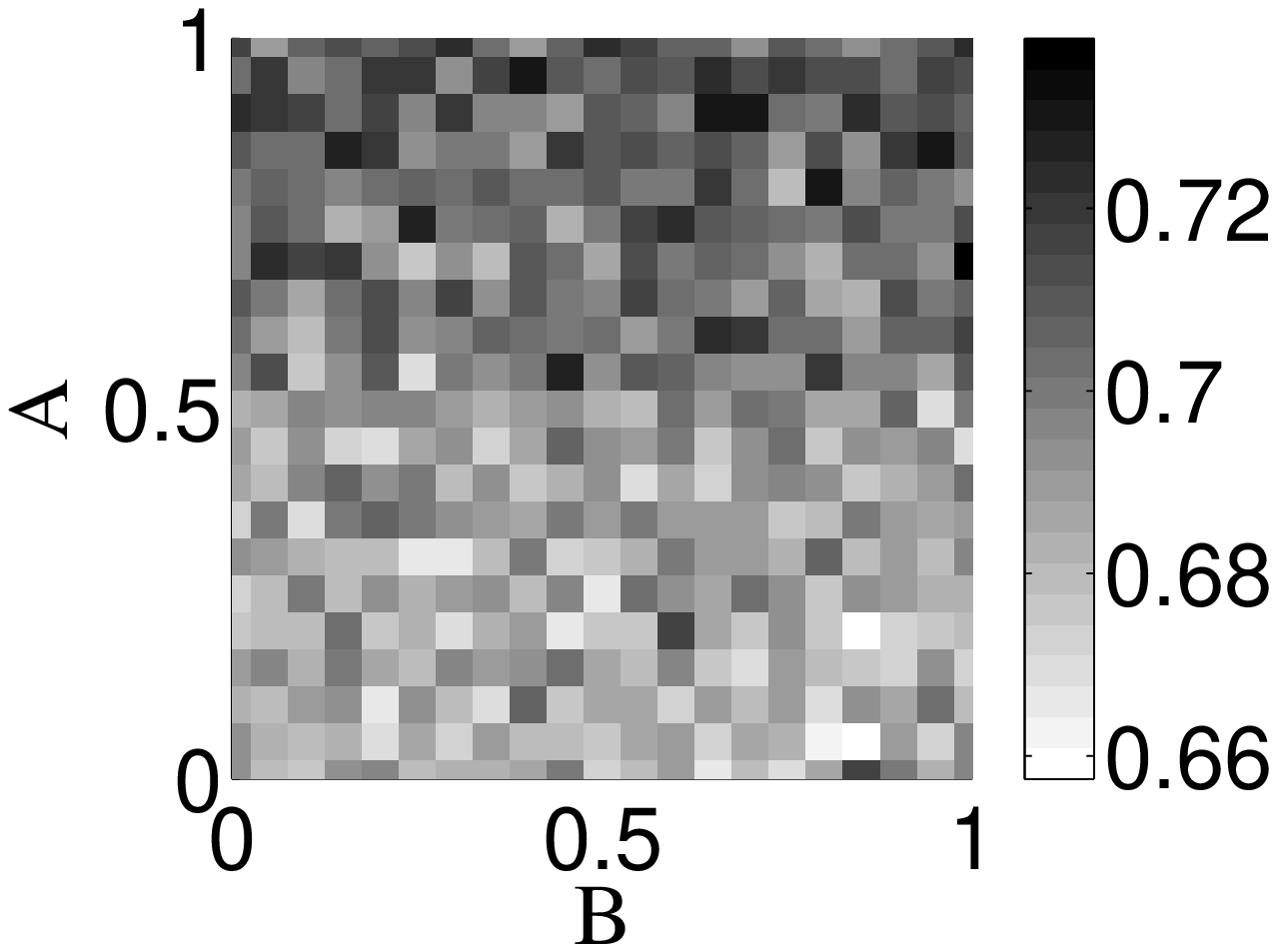} &
\includegraphics[scale=0.34]{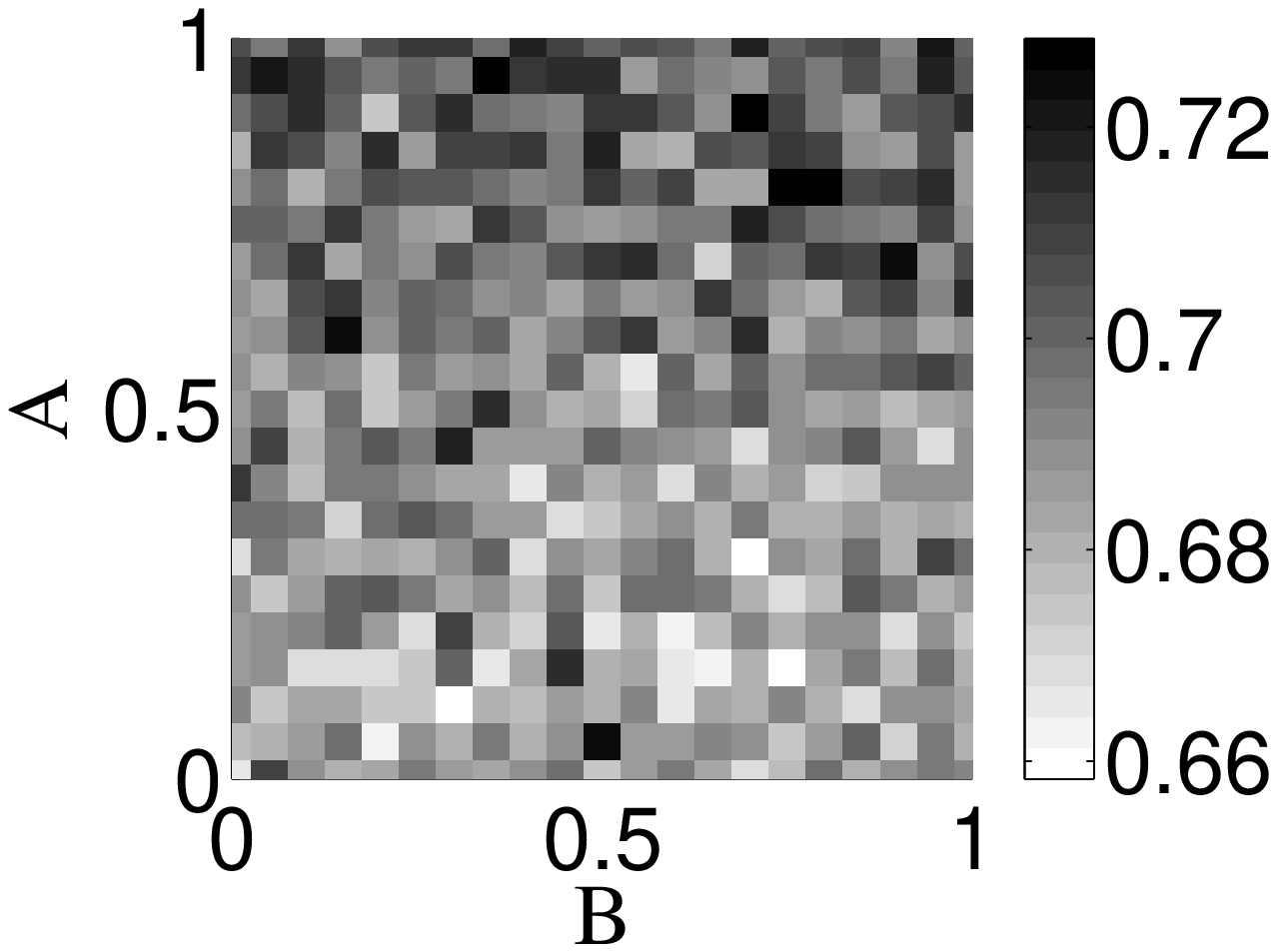} \\
(c) & (d) \\
\end{tabular}
\caption{Leaning, $\lambda_1$, of Eqn.\ \ref{eqn:nonlinearEX} computed $\delta_y=C$ as function of all three unitless parameters in Eqn.\ \ref{eqn:nonlinearEX}, $A$, $B$, and $C$.  The leaning agrees with intuition in this example for all the tested parameter values.  The subplots are each for different values of $C$, (a) $C=0.2$, (b) $C=0.4$, (c) $C=0.6$, and (d) $C=0.8$.}
\label{fig:nonlin1}
\end{figure}

Figure \ref{fig:nonlin1} shows the weighted mean observed leaning using the 1-standard assignment, i.e.,\ $\lambda_1$, agrees with intuition over the considered domains of $A$, $B$, and $C$ if the tolerance domain for $\mathbf{Y}$ is set to the noise level, i.e.,\ $\delta_y = C$.  The result of $\mathbf{X}\xrightarrow{lean}\mathbf{Y}$ shows that causal inference using leanings on data sets generated from nonlinear dynamics can be performed similarly, and can lead to similarly intuitive results, as the data sets generated from linear dynamics.

\subsection{Coupled Logistic Map Example}
\label{sec:2Pop}

Proponents of SSR time series causality tools have pointed out the limitations of tools like lagged cross-correlation and Granger causality when the dynamics exhibit chaotic behavior \cite{Sugihara2012}.  A chaotic system is considered in this section.

Consider the nonlinear dynamical system of
\begin{eqnarray}
\label{eqn:2pop}
\left\{\mathbf{X},\mathbf{Y}\right\} = \left\{\{x_t\},\{y_t\}\right\}
\end{eqnarray}
where $t=0,1,2,\ldots,L$,
\begin{equation*}
x_t = x_{t-1}\left(r_x-r_x x_{t-1}-\beta_{xy} y_{t-1}\right)
\end{equation*}
and
\begin{equation*}
y_t = y_{t-1}\left(r_y-r_y y_{t-1}-\beta_{yx} x_{t-1}\right)
\end{equation*}
where the parameters $r_x,r_y,\beta_{xy},\beta_{yx}\in\mathbb{R}\ge 0$.  This pair of equations is a specific form of the two-dimensional coupled logistic map system often used to model population dynamics \cite{Lloyd1995} and it was a system used in in the introduction of cross convergent mapping (CCM) which is a SSR time series causality tool \cite{Sugihara2012}.

Sugihara et al.\ \cite{Sugihara2012} note that $\beta_{xy}>\beta_{yx}$ intuitively implies $\mathbf{Y}$ ``drives'' $\mathbf{X}$ more than $\mathbf{X}$ ``drives'' $\mathbf{Y}$, and vice versa.  Such intuition, however, can be difficult to justify for all instances of Eqn.\ \ref{eqn:2pop}.  The $x_{t-1}$ term that appears in $y_t$ can be seen as a function of $x_{t-2}$ with coefficients of $\beta_{yx}r_x$.  These product coefficients suggest that if $r_x>r_y$, then $\mathbf{X}$ may be seen as the stronger driver in the system even if $\beta_{yx}<\beta_{xy}$.  The same argument can be made, with the appropriate substitutions, to show that $\mathbf{Y}$ may be seen as the stronger driver in the system even if $\beta_{xy}<\beta_{yx}$.  As such, there is no clear intuitive causal inference for this system.  The conjectures presented in this paragraph, however, are supported by the leaning calculations (using the 1-standard assignment).

\begin{figure}[ht]
\begin{tabular}{cc}
\includegraphics[scale=0.34]{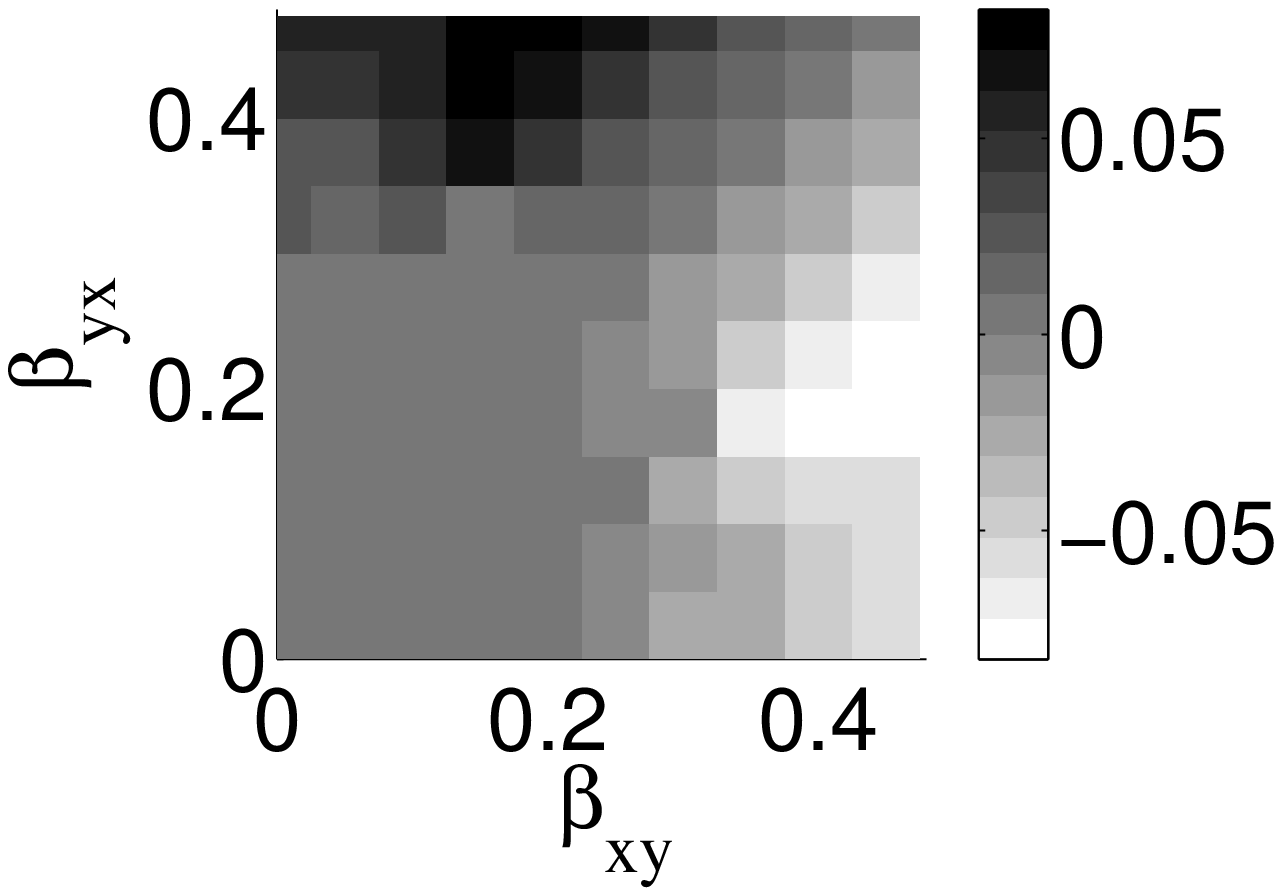} &
\includegraphics[scale=0.34]{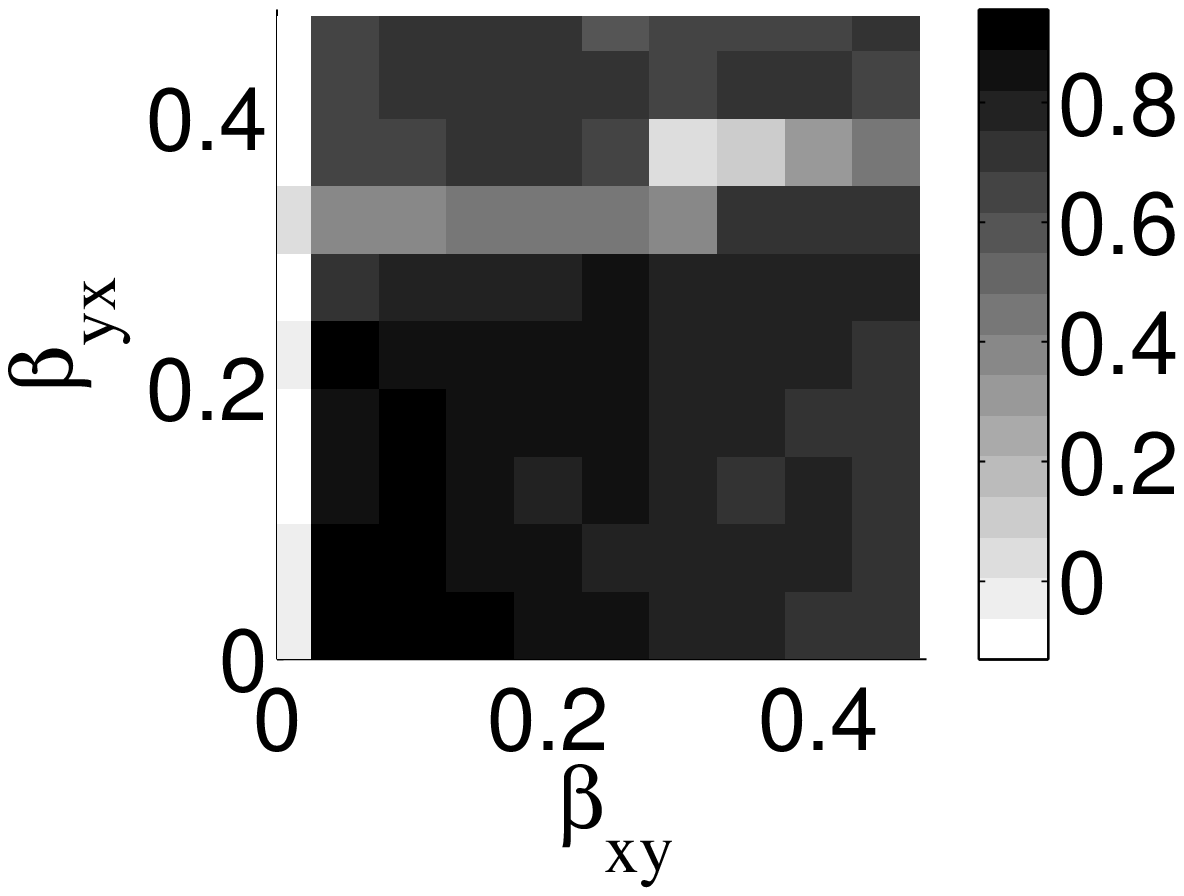} \\
(a) & (b) \\
\includegraphics[scale=0.34]{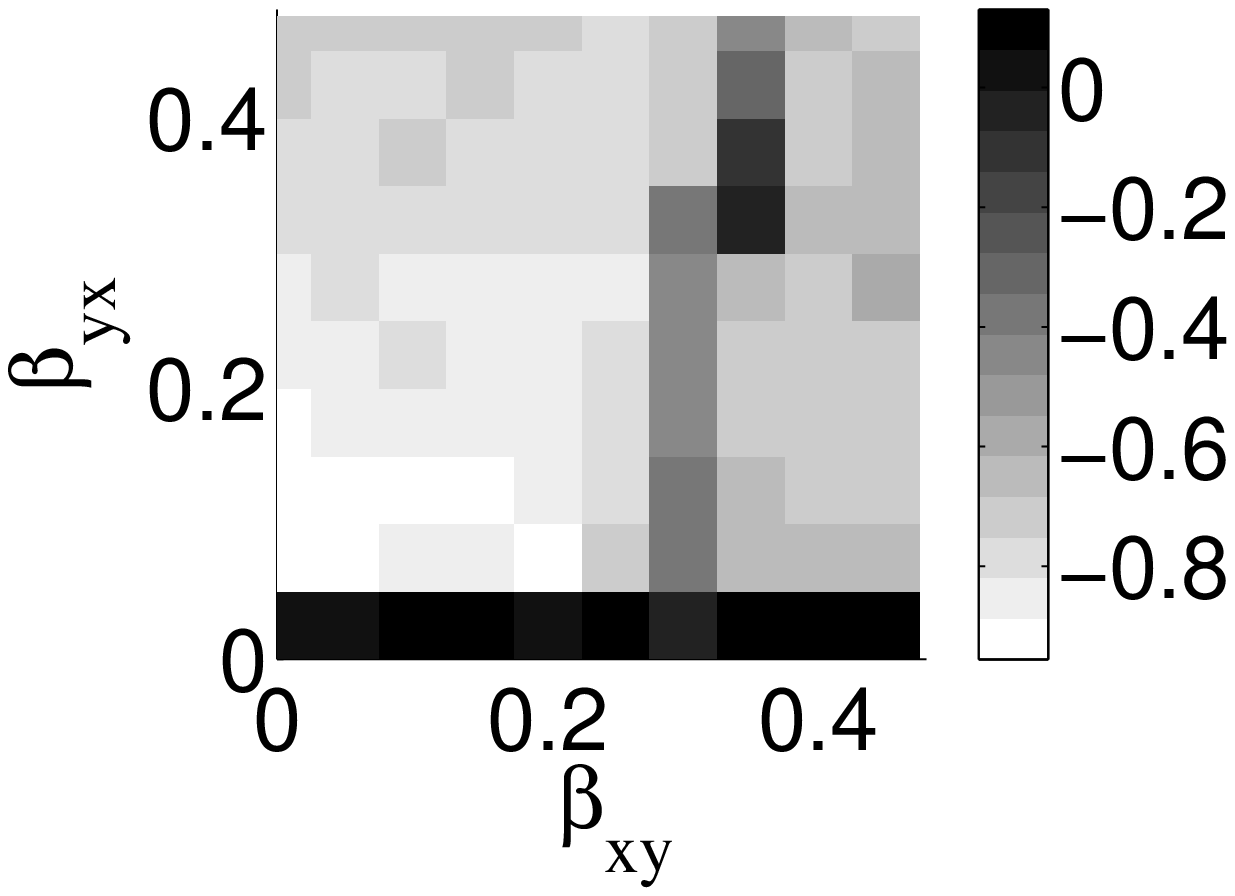} &
\includegraphics[scale=0.34]{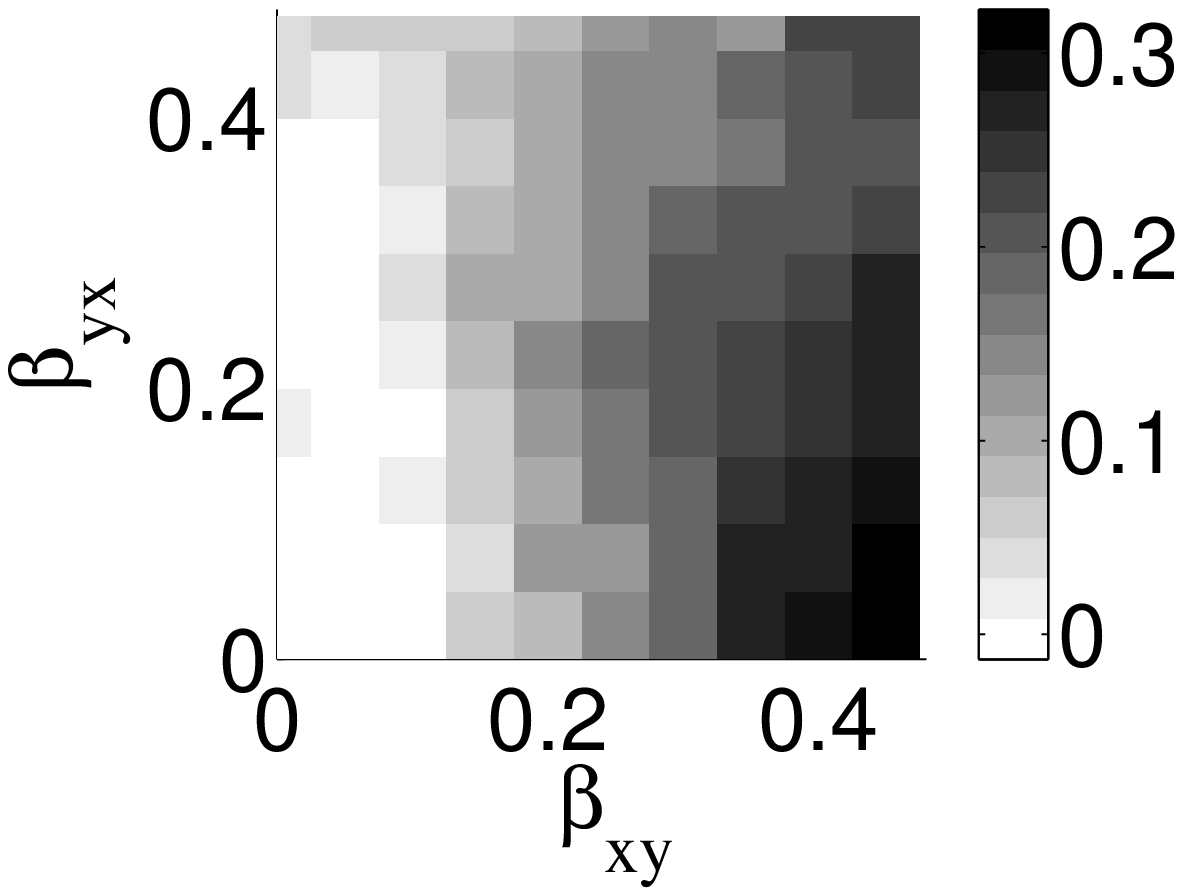} \\
(c) & (d) \\
\end{tabular}
\caption{Leaning, $\lambda_1$, of Eqn.\ \ref{eqn:2pop} is a function of four unitless parameters in Eqn.\ \ref{eqn:2pop}, $r_x$, $r_y$, $\beta_{xy}$, and $\beta_{yx}$ (along with the initial conditions $x_0$ and $y_0$, which are fixed in this example).  The tolerance domains are set as $\delta_x = \sigma_{x_t-\langle x_t \rangle}$ and $\delta_y = \sigma_{y_t-\langle y_t \rangle}$.  The leaning is defined using the 1-standard assignment, so $\lambda_1>0\Rightarrow\mathbf{X}\xrightarrow{lean}\mathbf{Y}$.  The subplots are each for different parameter values, (a) $r_x = r_y = 3.5$, (b) $r_x = 4.0$, $r_y = 2.0$, (c) $r_x = 2.0$, $r_y = 4.0$, and (d) $r_x = 3.8$, $r_y = 3.2$.}
\label{fig:2pop}
\end{figure}
Figure \ref{fig:2pop} shows four instances of Eqn.\ \ref{eqn:2pop} with different values for $r_x$ and $r_y$.  Each instance has a library length of $L=500$ and initial conditions of $x_0 = 0.4$ and $y_0 = 0.4$.  There is no clear, intuitive driver in this example, so both tolerance domains must be set in the leaning calculation.  The leaning is calculated using the 1-standard cause-effect assignment and estimated tolerance domains of $\delta_x = \sigma_{x_t-\langle x_t \rangle}$ and $\delta_y = \sigma_{y_t-\langle y_t \rangle}$.

Figure \ref{fig:2pop}(a) shows the intuition of $\beta_{xy}<\beta_{yx}\Rightarrow\mathbf{X}\xrightarrow{lean}\mathbf{Y}$ can be true if $r_x=r_y$.  However, Figure \ref{fig:2pop}(b) and (c) shows $r_x>r_y\Rightarrow\mathbf{X}\xrightarrow{lean}\mathbf{Y}$ and $r_x<r_y\Rightarrow\mathbf{Y}\xrightarrow{lean}\mathbf{X}$ can be strong enough implications to make the values of $\beta_{xy}$ and $\beta_{yx}$ irrelevant over the considered domains.  Figure \ref{fig:2pop}(d) shows this effect can be pronounced even in instances of Eqn.\ \ref{eqn:2pop} where $r_x$ and $r_y$ are close.  

Sugihara et al.\ \cite{Sugihara2012} also discuss how a naive \footnote{Sugihara et al.\ \cite{Sugihara2012} do not explore any of the numerous non-linear extenstions of Granger causality.  The theoretical foundations of Granger causality are independent of its practical implementations, and failures of Granger causality may be failures of a specific implementation, e.g., using linear forecast models with non-linear data \cite{Granger1980}.} application of Granger causality to the system described in Eqn.\ \ref{eqn:2pop} may lead to conclusions that do not agree with intuition, while CCM does.  The causal inference suggested by the leaning calculations of this subsection implies both CCM and leanings may be useful time series causality tools in situations where Granger causality is not.  It has also been shown that CCM may fail to agree with intuition in example systems for which it has already been shown that leaning calculation do, e.g., Section \ref{sec:rlcirc} \cite{Weigel2014}.

The complexity of determining causal relationships in this system may make the system less of a convincing example of the leaning calculation than the previous examples.  However, Figure \ref{fig:2pop} shows the weighted mean observed leaning using the 1-standard cause-effect assignment can provide causal inferences that may be considered intuitively justifiable, even if the system does not have an unequivocal driver. 

\subsection{Impulse with Multiple Noisy Responses Example}

All of the previous examples have ignored possible causal confounders.  The presence of confounders in the system is a serious problem for causal inference in general \cite{Rubin2015,Pearl2000}.  Time series causality usually seeks to answer the less general causal inference question of ``Given two times series, which may be considered the stronger driver?'' Nevertheless, some bivariate time series causality tools consider causal inference in systems with potential confouders by trying to relate the estimated bivariate driving relationships within a collection of more than two time series data sets (e.g., see CCM \cite{Sugihara2012}).  The next example explores the use of leaning calculations in such a scenario.

Consider the multivariate system of
\begin{eqnarray}
\label{eqn:3var}
\bar{\mathbf{\tau}}_L = \left\{\mathbf{X},\mathbf{Y},\mathbf{Z}\right\} = \left\{\{x_t\},\{y_t\},\{z_t\}\right\}
\end{eqnarray}
where $t=0,1,2,\ldots,L$,
\begin{equation*}
x_t = \left\{
  \begin{array}{lr}
    2 & t = 1\\
    0 & \forall\; t\in\{t\;|\;t\neq 1 \;\mathrm{and}\; t\bmod 5 \neq 0\}\\
    2 & \forall\; t\in\{t\;|\;t\bmod 5 = 0\}
  \end{array}
\right.
\end{equation*}
and
\begin{equation*}
y_t = x_{t-1} + B\eta_t\;\;,
\end{equation*}
and either (case 1)
\begin{equation}
z_t = y_{t-1}
\end{equation}
or (case 2)
\begin{equation}
z_t^\prime = y_{t-1} + y_t = y_{t-1} + x_{t-1} + B\eta_t
\end{equation}
or (case 3)
\begin{equation}
z_t^{\prime\prime} = y_{t-1} + x_{t-1} + z_{t-1}
\end{equation}
with $y_0 = 0$, $B\in\mathbb{R}\ge 0$, $\eta_t\sim\mathcal{N}\left(0,1\right)$, and $L=500$.

In case 1, $\mathbf{Z}$ depends directly on $\mathbf{Y}$ and indirectly on $\mathbf{X}$ (through $\mathbf{Y}$, which depends directly on $\mathbf{X}$).  The intuitive causal inference is then $\mathbf{Y}\xrightarrow{lean}\mathbf{Z}$ and $\mathbf{X}\xrightarrow{lean}\mathbf{Z}$.  Case 2, despite the additional $\mathbf{Y}$ dependence in $\mathbf{Z}$, has the same intuitive causal inference as case 0.  In case 2, $\mathbf{Z}$ depends directly on itself and both $\mathbf{Y}$ and $\mathbf{X}$.  Case 3 also has the same intuitive causal inference.

\begin{figure}[ht]
\begin{tabular}{c}
\includegraphics[scale=0.45]{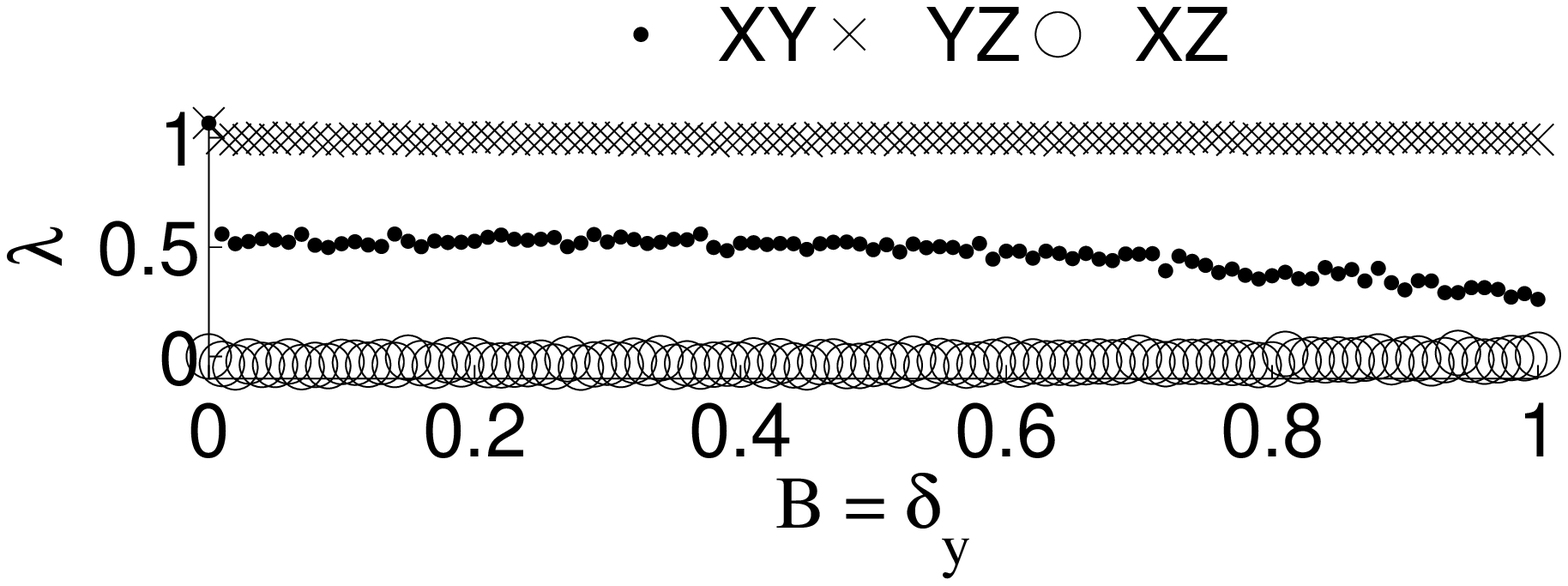} \\ 
(a) \\
\includegraphics[scale=0.45]{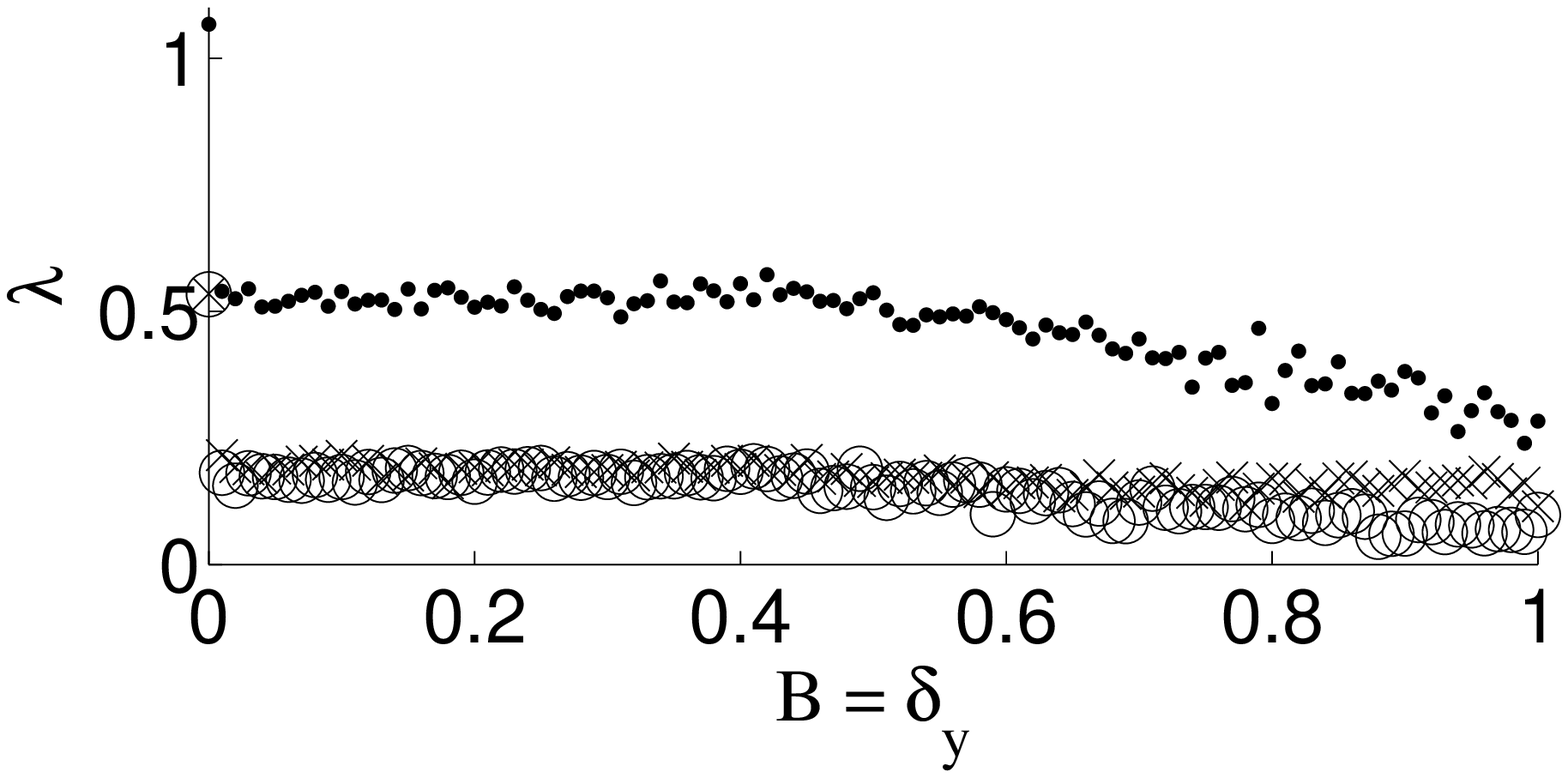} \\
(b) \\
\includegraphics[scale=0.45]{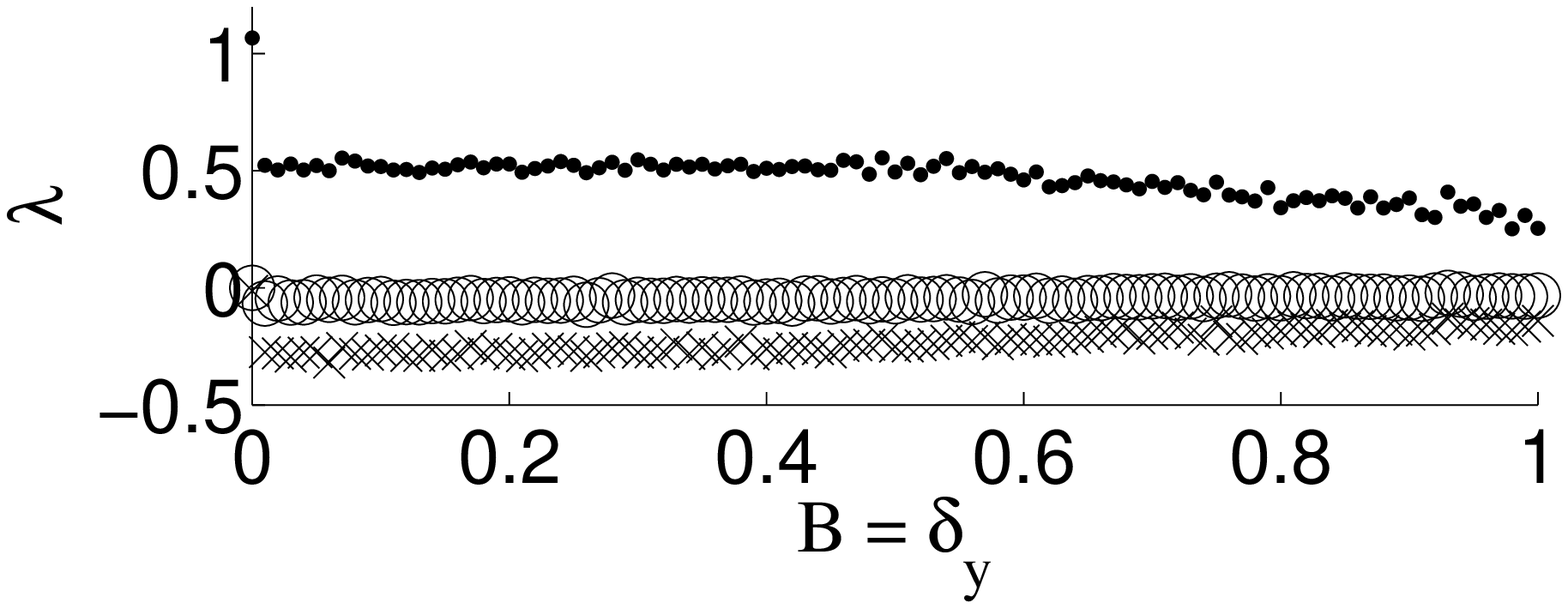} \\
(c) \\
\end{tabular}
\caption{The (unitless) weighted mean observed leaning using the 1-standard cause-effect assignment, $\lambda_1$ only leads to causal inferences that agree with intuition for all points within the considered noise level domain, $B$, in this example for case 2.  $B$ is the unitless noise parameter found in Eqn.\ \ref{eqn:3var}, and the symmetric tolerance domains for $\mathbf{Y}$ and $\mathbf{Z}$ are set to this value.  All of the leanings are expected to be positive in each case.  The subplots are each for different the three different cases, (a) $\mathbf{Z} = \{z_t\}$ (case 1), (b) $\mathbf{Z} = \{z^\prime_t\}$ (case 2), and (c) $\mathbf{Z} = \{z^{\prime\prime}_t\}$ (case 3).}
\label{fig:3var}
\end{figure}
Figure \ref{fig:3var} shows the weighted mean observed leaning using the 1-standard cause-effect assignment (with $\delta_x = 0$ and $\delta_y = \delta_z = B$), $\lambda_1$, may lead to causal inferences that do not agree with intuition for case 1 and case 3, even though case 2 agrees with intuition for all points within the considered noise levels domain.  For case 2, the leaning calculation implies $\mathbf{X}\xrightarrow{lean}\mathbf{Y}$ and $\mathbf{Y}\xrightarrow{lean}\mathbf{Z}$ as expected, but it also seem to imply that no causal inference can be made about the relationship between $\mathbf{X}$ and $\mathbf{Z}$.  For case 3, the leaning calculation also implies no causal inference can be made about the relationship between $\mathbf{X}$ and $\mathbf{Z}$, but, unlike case 1, it also implies $\mathbf{Z}\xrightarrow{lean}\mathbf{Y}$, which is counter-intuitive.

These results may imply that $\lambda_1$ is unable to identify confounded driving (i.e., situations in which the effect of the driving variable is mediated by another variable).  For example, in case 1, the driving of $\mathbf{Z}$ by $\mathbf{X}$ occurs through the interaction of $\mathbf{Y}$ and $\mathbf{Z}$.  For case 1, $\lambda_1$ implies $\mathbf{X}\xrightarrow{lean}\mathbf{Y}\xrightarrow{lean}\mathbf{Z}$ but does not imply $\mathbf{X}\xrightarrow{lean}\mathbf{Z}$.  For case 3, $\lambda_1$ implies $\mathbf{X}\xrightarrow{lean}\mathbf{Y}\xleftarrow{lean}\mathbf{Z}$, which may imply that $\lambda_1$ is not a reliable causal inference tool in autoregressive systems.

The results of Figure \ref{fig:3var} may also be considered an indication that the cause-effect assignment is insufficient.  It was previously mentioned that exploratory causal analysis using the leaning would involve comparing several different cause-effect assignments.  The set of tested cause-effect assignments need not only include $l$-standard assignments.  Consider the weighted mean observed leaning, $\lambda_{AR}^{xy}$ using the 1-AR cause-effect assignment, i.e., $\{C,E\} = \{x_{t-1}\;\mathrm{ and }\;y_{t-1},y_t\}$.  Table \ref{tab:3var} shows this leaning calculation, using $\delta_y=\delta_z=B=0.6$, for the same bivariate relationships shown in Figure \ref{fig:3var}.
\begin{table}
\begin{tabular}{lccc}
 & case 1 & case 2 & case 3\\
 \hline
 $\lambda_{AR}^{xy}$ & 0.150 & 0.159 & 0.169\\
 $\lambda_{AR}^{yz}$ & -0.002 & 0.133 & 0.447\\
 $\lambda_{AR}^{xz}$ & 0.691 & 0.030 & 0.735\\
\end{tabular}
\caption{The leaning calculation depends strongly on the cause-effect assignment.  The table shows the weighted mean observed leaning using the 1-AR assignment and may be compared with Figure \ref{fig:3var}, which showed this leaning calculation using the 1-standard assignment.}
\label{tab:3var}
\end{table}

Table \ref{tab:3var} implies $\mathbf{X}\xrightarrow{lean}\mathbf{Y}\xrightarrow{lean}\mathbf{Z}$ for case 2 and 3, as expected, but not for case 1 (which $\lambda_1$ did imply).  The leaning calculations are part of an exploratory causal analysis and must be considered using several different cause-effect assignments when trying to understand the potential causal structure of a set of times series data.  The cause-effect assignments can also be expanded beyond the bivariate and autoregressive definitions, e.g., $\{C,E\} = \{x_{t-1}\;\mathrm{ and }\;y_{t-1}\;\mathrm{ and }\;z_{t-1},y_t\}$, but such extensions will not be considered in this article.

\section{Empirical Data}
\label{sec:emp}
Empirical data sets with known (or assumed) causal relationships may be used to understand how exploratory causal inference using leanings might be done if the system dynamics are unknown (or sufficiently complicated to make first principle numerical comparisons cumbersome).  The examples shown in this section are intended to demonstrate that causal inference using leaning calculations can agree with the causal ``truth'' in empirical data sets.  The analysis shown here is not expected to illustrate how the leaning may be used for exploratory causal analysis of empirical data for which there is no causal ``truth''.  Such analysis is expected to be more complicated than that which is shown here (e.g., involving multiple tolerance domain calculations and the comparison of different cause-effect assignments).

\begin{figure}[ht]
\begin{tabular}{c}
\includegraphics[scale=0.46]{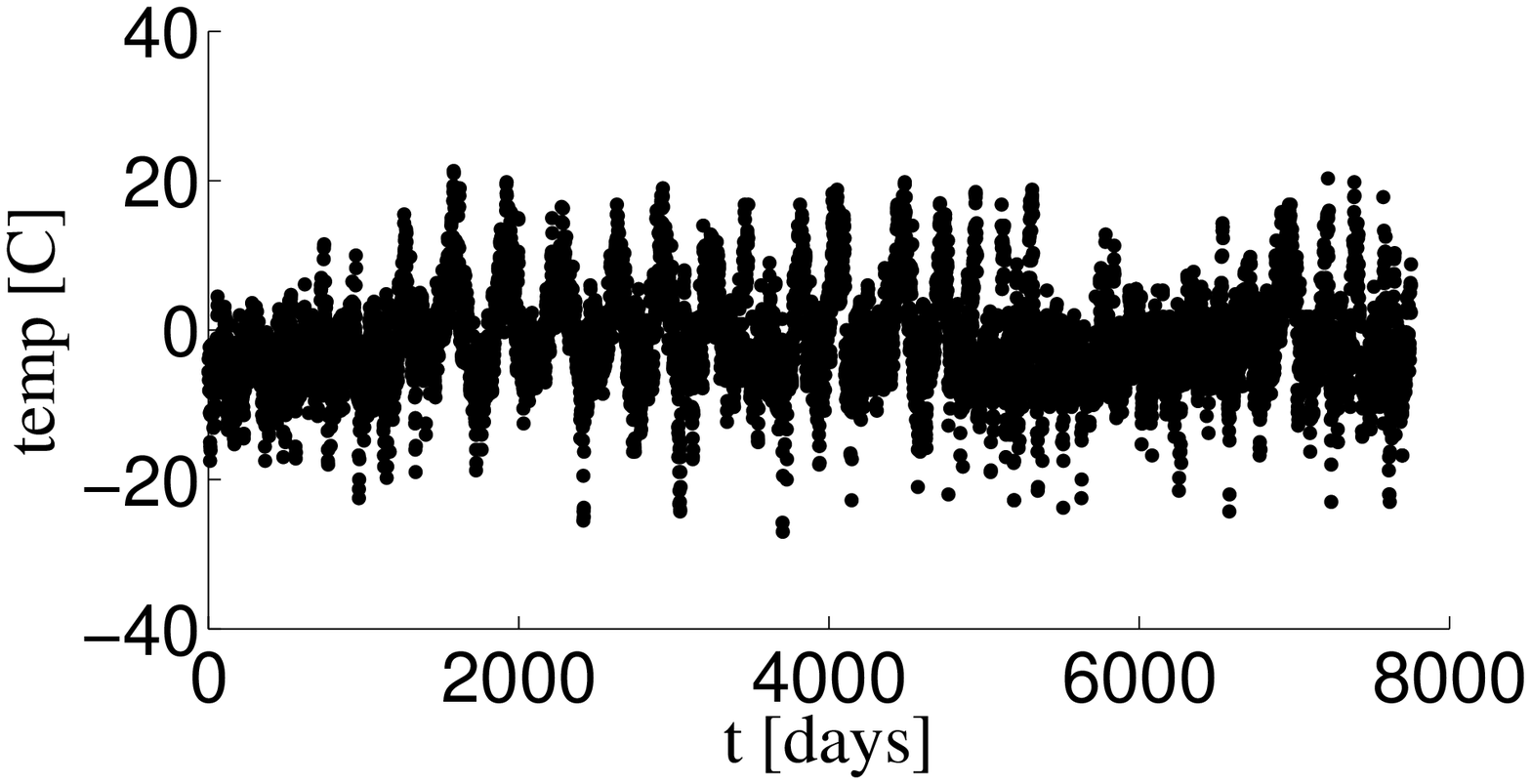} \\
(a) \\ 
\includegraphics[scale=0.46]{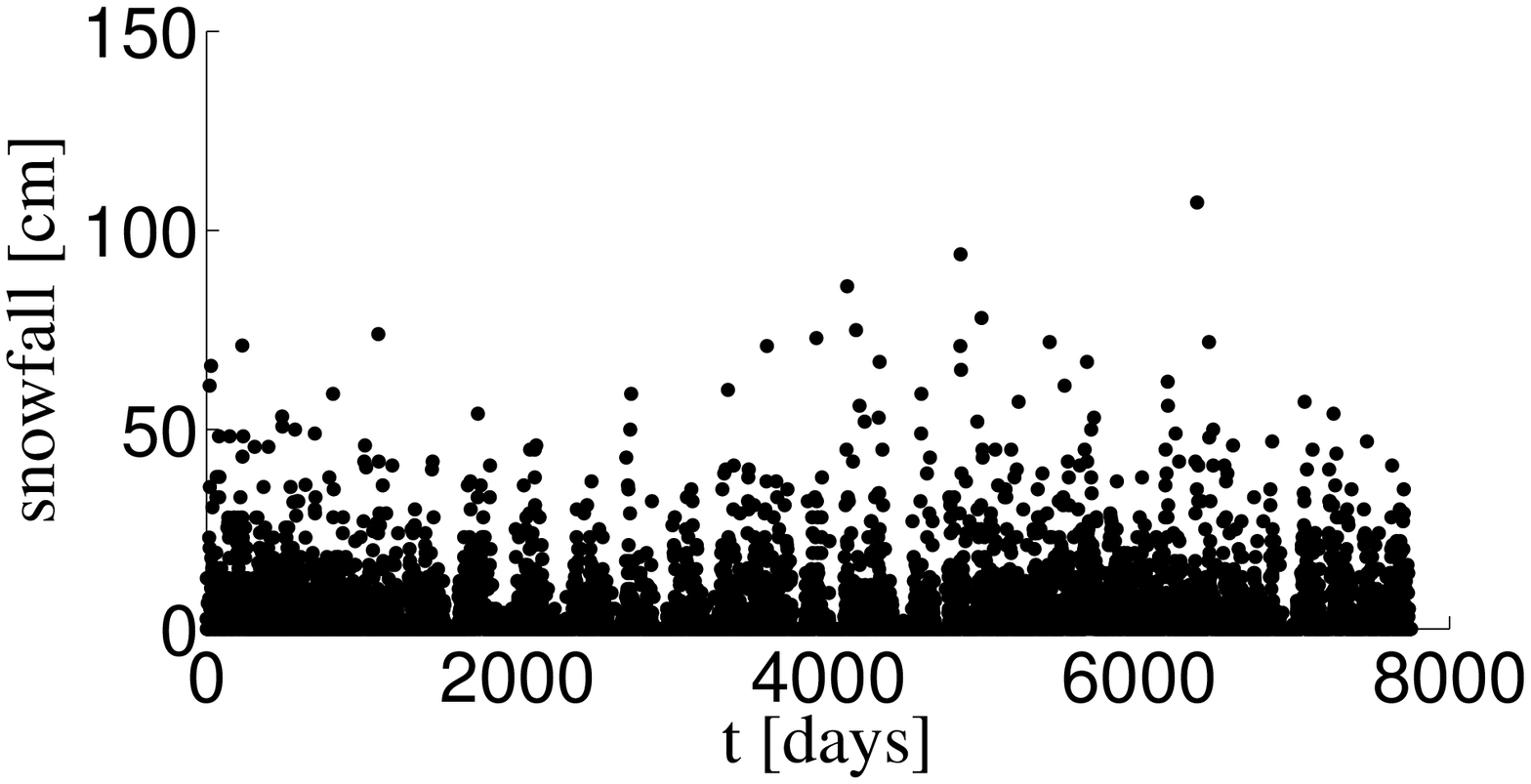} \\
(b) \\
\end{tabular}
\caption{This times series data pair is expected to have the causal relationship of $X\rightarrow Y$, where $X$ or $Y$ is marked in parenthesis for each time series.  Subplot (a) is the mean temperature (X) and (b) is the snow fall (Y).}
\label{fig:empdata}
\end{figure}

Figure \ref{fig:empdata} shows a time series pair with causal ``truth'' from the UCI Machine Learning Repository (MLR) \cite{bache2013}.  This data repository is a collection of data sets (some of which are time series) with known, intuitive, or assumed causal relationships meant for use in the testing of causal discovery algorithms in machine learning \cite{bache2013}.  

Figure \ref{fig:empdata}(a) and (b)  are times series of the daily snowfall (the expected response) and mean temperature (the expected driver) from July 1 1972 to December 31 2009 at Whistler, BC, Canada (Latitude: 50$^\circ$04$^\prime$04.000$^{\prime\prime}$ N, Longitude: 122$^\circ$56$^\prime$50.000$^{\prime\prime}$ W, Elevation: 1835.00 meters).  From \cite{bache2013}, ``Common sense tells us that X [mean temperature] causes Y [snow fall] (with maybe very small feedback of Y on X). Confounders are present (e.g., day of the year).''  These time series correspond to data set 87 of the MLR \cite{bache2013}.

As noted previously, the primary difficulty in using the leaning for exploratory causal analysis is the determination of the cause-effect assignment and tolerance domains.  The above data is meant only to illustrate the use of leanings, so while a thorough analysis of the noise in the system should precede the leaning calculations, such a step is avoided here for brevity.  

The symmetric tolerance domains are estimated using the maximum standard deviations of the $n$ sets of binned points of an $n$-bin histogram of the normalized time series data $\mathbf{X}^\prime$ and $\mathbf{Y}^\prime$, where $\mathbf{X}^\prime = \frac{\mathbf{X}-\langle \mathbf{X} \rangle}{\sigma_\mathbf{X}}$, $\mathbf{Y}^\prime = \frac{\mathbf{Y}-\langle \mathbf{Y} \rangle}{\sigma_\mathbf{Y}}$, and $n=\lfloor 0.1L\rfloor$ (i.e., $n$ is the closest integer that is not larger than 10\% of the library length $L$).  This estimation is similar to the {\em n-bin mean standard deviation} technique discussed in Sec.\ \ref{sec:IR}.

The cause-effect assignment will be set naively because, again, the purpose of this article in not to study these particular time series in detail.  To reiterate the previous comment regarding tolerance domains, detailed study would be required to have confidence in using leanings for exploratory causal analysis.  However, the convenience of having causal ``truths'' is that we can take the naive approach of simply testing many different cause-effect assignments and compare the results to the expected causal inference.

\begin{figure}
\includegraphics[scale=0.43]{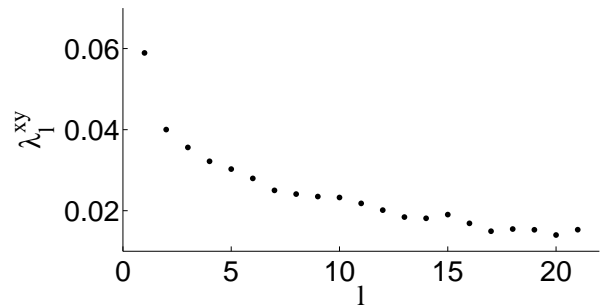}
\caption{The weighted mean observed leaning, $\lambda_l^{xy}$, using the $1$-standard cause-effect assignment for $l\in[0,21]$.  The time series pair $\mathbf{\tau}_1 = (\mathbf{X},\mathbf{Y})$ with $\mathbf{X}$ and $\mathbf{Y}$ shown in Figure \ref{fig:empdata}(a) and (b), respectively.  The expected causal inference is $\mathbf{X}\xrightarrow{lean}\mathbf{Y}$; i.e., the expectation is $\lambda_l^{xy} > 0$ for every point in the plotted domain.}
\label{fig:emp}
\end{figure}

Figure \ref{fig:emp} shows the weighted mean observed leaning, $\lambda_l^{xy}$, using the $1$-standard cause-effect assignment with $l\in[0,21]$ and using $\delta_x$ and $\delta_y$ estimated in the manner described above.    

Figure \ref{fig:emp} shows the leanings, using the given tolerance domains, imply causal inferences that agree with the causal truths for the tested pair with $1$-standard assignments.    

This example also highlights the problem of determining which $l$-standard assignment to use for the causal inference.  If it is decided that the causal inference depends on
\begin{equation}
\lambda^{xy}_{max} = \lambda^{xy}_{l^\prime}
\end{equation}
where
\begin{equation}
|\lambda^{xy}_{l^\prime}| = \max_l |\lambda^{xy}_l|\;\;,
\end{equation}
then $\lambda^{xy}_{max}=0.040\Rightarrow\mathbf{X}\xrightarrow{lean}\mathbf{Y}$, which agrees with the causal truth.   

The NASA OMNI data set consists of hourly-averaged time series measurements of several different space weather parameters from 1963 to present, collected from more than twenty different satellites, along with sunspot number and several different geomagnetic indices, including $D_{st}$, collected from the NOAA National Geophysical Data Center \cite{King2005}.  The disturbance storm time, $D_{st}$, is a measure of geomagnetic activity \cite{IAGA}.  The magnetic field measurements in the OMNI data sets, specifically $B_z$ in GSE coordinates \cite{Hapgood1992}, is believed to be a driver of $D_{st}$ \cite{Gonz1994}.

Let $\mathbf{P_{z}^L}=\{\{B_z(t^\prime)\},\{D_{st}(t^\prime)\}\;|\;t^\prime\in[t^\prime_0,L]\}$ be an ordered subset of the available time series data $\{\{B_z(t)\},\{D_{st}(t)\}\;|\;t=0,1,2,\ldots,N\}$ where $N$ is the number of hourly data points in the OMNI data set.  If $\lambda_1^{z}$ is the weighted mean observed leaning for $\mathbf{P_{z}^L}$ using the 1-standard cause-effect assignment, then $n$ samples of $\mathbf{P_{z}^L}$, each with a different $t^\prime_0$,  would produce a set of $n$ leanings, $\{\lambda_1^{z}\}$, from which the causal inference could be drawn.

\begin{figure}[ht]
\begin{tabular}{c}
\includegraphics[scale=0.45]{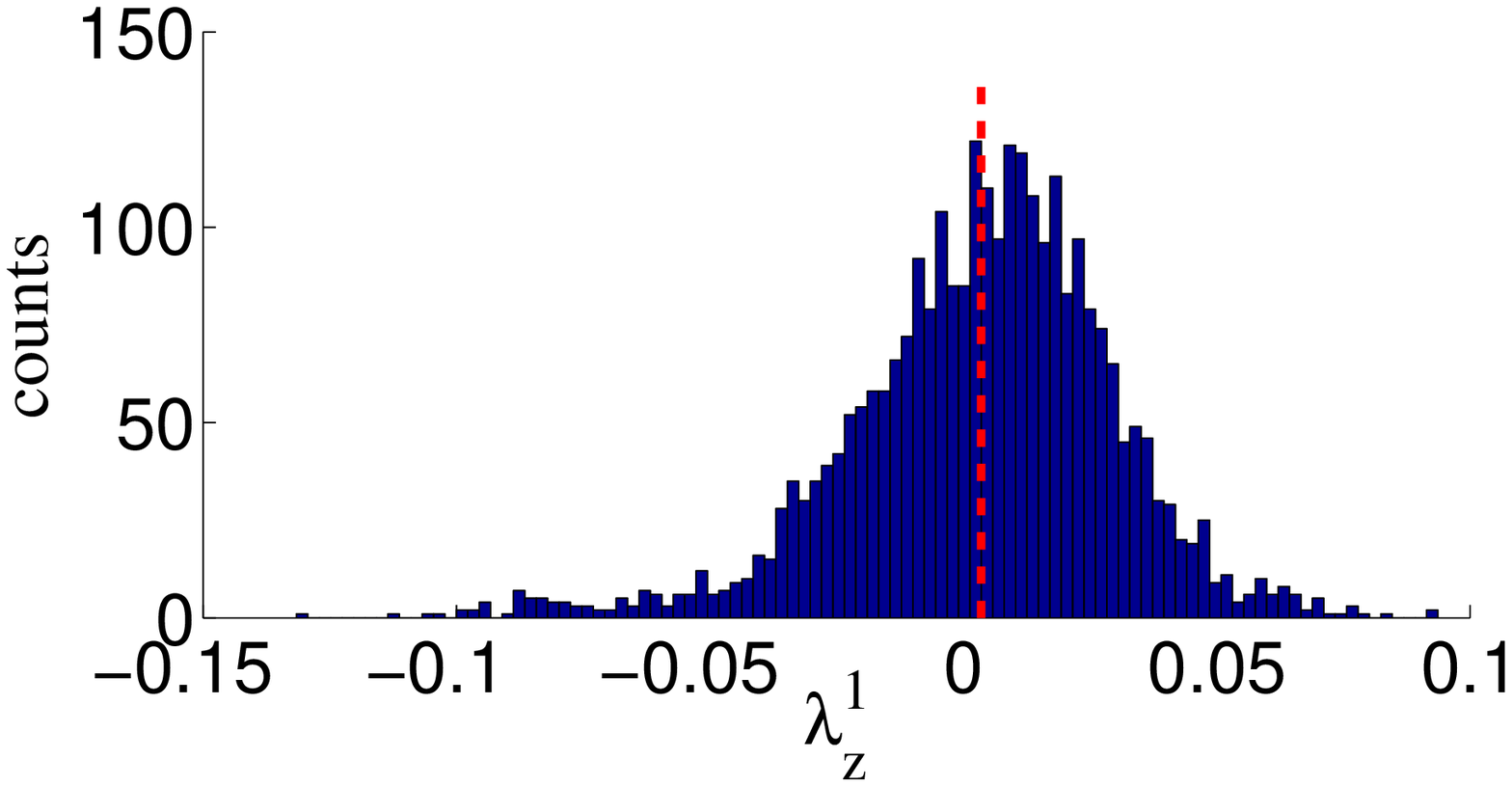} \\ 
(a) \\
\includegraphics[scale=0.45]{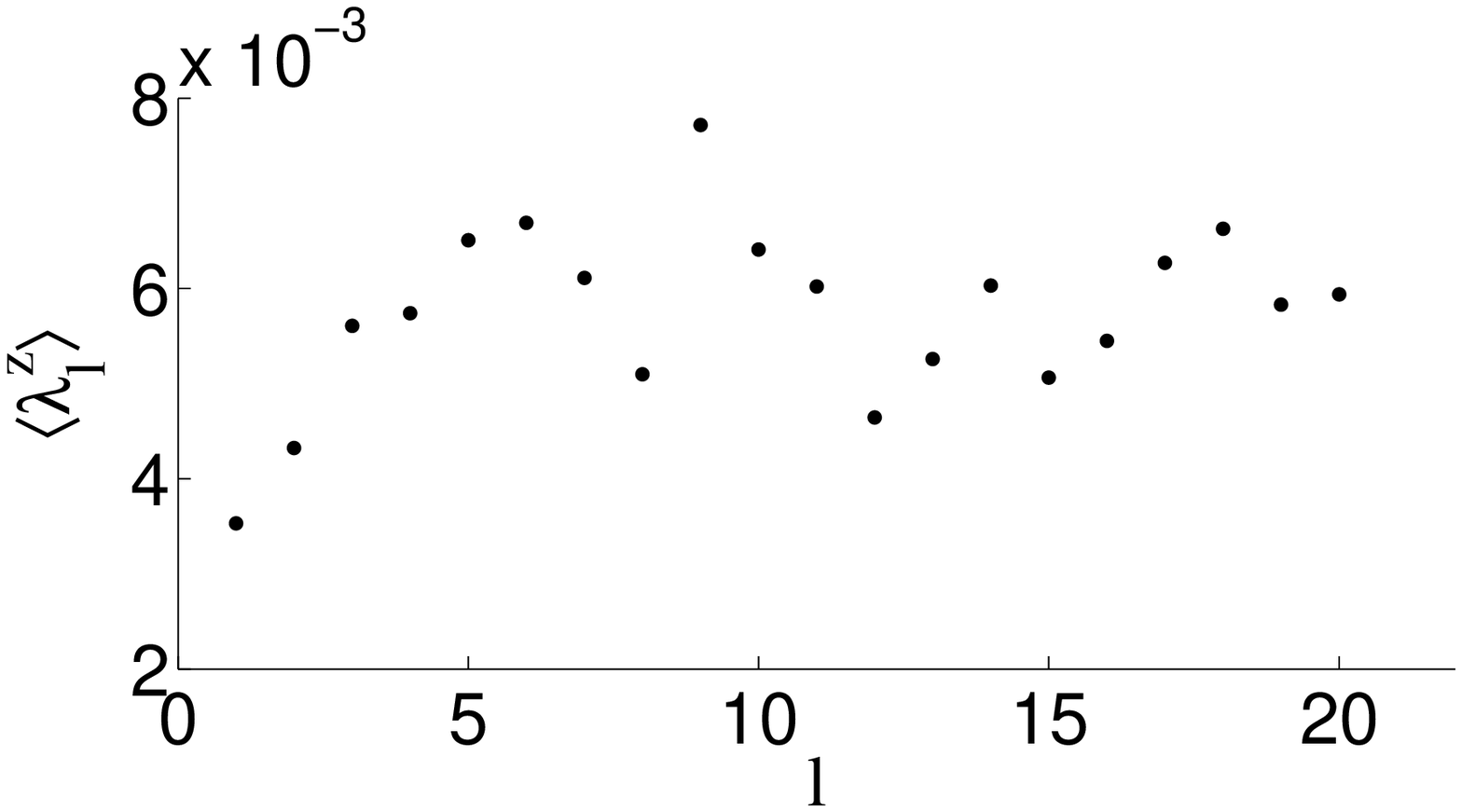} \\ 
(b) 
\end{tabular}
\caption{(Color available online) (a) A histogram of the $(B_z^\prime,D_{st})$ set of $n=10^4$ (unitless) weighted mean observed leanings, using the 1-standard cause-effect assignment, i.e., $\lambda^1_{z}$, show causal inferences that agree with intuition for the OMNI data set.  The red dashed lines show the algebraic mean of the sets.  The geomagnetic field component $B_z^\prime$ is calculated by taking the $B_z$ times series in the OMNI data set and then setting $B_z^\prime = 0$ if $B_z>0$ \cite{Burton1975,Dungey1961}. (b) The algebraic means of the aforementioned sets of weighted mean observed leanings, i.e., $\langle\lambda^z_l\rangle$, are positive for all leanings calculated using $l$-standard assignment given $l\in[1,20]$.}
\label{fig:dsthist}
\end{figure}

Let $L=500$ and $n=10^4$.  The symmetric tolerance domains are naively set with $f\sigma_{|B_{z}^\prime-\langle B_{z}^\prime\rangle|}$ and $f\sigma_{|D_{st}-\langle D_{st}\rangle|}$ for each sampled times series of length $L$ with f=0.05.  The starting points for each time series are sampled from a uniform distribution over $[0,N-L]$.  Figure \ref{fig:dsthist}(a) shows the causal inference drawn from each set of leanings agrees with intuition, i.e., $B_{z}\xrightarrow{lean}D_{st}$, if the causal inference is based on, e.g., the mean value from the set of leanings $\langle\lambda^z_l\rangle$ with $l=1$.  The algebraic means $\langle\lambda^z_l\rangle$ found using different $l$-standard assignments with $l\in[1,20]$ are shown in Figure \ref{fig:dsthist}(b).  The causal inference is $B_{z}\xrightarrow{lean}D_{st}$ for every $l$ in the plotted domain.

This example is the first for which a set of leanings has been used for causal inference, which may imply statistical testing should be used.  The sample mean was used for causal inference and happened to agree with intuition for this example, but would the same conclusion be drawn using a formal hypothesis test?  How should the null hypothesis and test statistic be constructed?  Such questions can be subtle (see, e.g., \cite{Rubin2015}).  The sampling procedure used to produce Figure \ref{fig:dsthist}(a) produces 2,796 defined leanings, 95\% of which are below $4.2\times 10^{-3}$ and 5\% of which are below $-4.0\times 10^{-3}$.  A 90\% confidence that the leaning falls in the interval $[-4.0\times 10^{-3},4.2\times10^{-3}]$, however, is not a strong indication that the data supports the intuitive causal structure.  The mean of the set is $\mu = 3.5\times 10^{-3}$, and the variance is $\sigma^2 = 7.0\times 10^{-5}$.  If it is assumed that the leaning in this example is distributed as $\mathcal{N}(\mu,\sigma^2)$, then a 95\% confidence interval may be $[\mu-2\sigma,\mu+2\sigma]=[-4.9\times 10^{-3},5.6\times 10^{-3}]$, which, again, does not strongly support the intuitive causal inference for this example.  Approximately 40\% of the leanings in this example are negative, which may imply that there is only a 60\% confidence that this data supports the intuitive causal inference, given the tolerance domains and cause-effect assignments.  

Suppose a null hypothesis is defined as $\langle\lambda_1^z\rangle = 0$.  The standard error is $\mathrm{SE}=\sigma/\sqrt{n}= 5.0\times 10^{-4}$, from which the t-test statistic follows \cite{Casella2002} as $t=\mu/\mathrm{SE}=7.08$.  A two-tailed t-test (i.e., calculating the $p$-value with an alternative hypothesis of $\langle\lambda_1^z\rangle \neq 0$) returns a $p$-value of approximately zero \footnote{This calculation, and all t-test calculations discussed in the section were performed with the {\sc MATLAB} function {\em ttest}.}, which implies the null hypothesis should be rejected in favor of the alternative at any significance level.  A right-tailed t-test (i.e., the alternative hypothesis is $\langle\lambda_1^z\rangle > 0$) also returns a $p$-value of approximately zero.  A left-tailed t-test (i.e., the alternative hypothesis is $\langle\lambda_1^z\rangle < 0$) returns a $p$-value of approximately one, which implies the null hypothesis cannot be rejected in favor of the alternative at any significance level.  These hypothesis tests seem to imply the population mean of the sampled leanings calculated in this example is likely not zero (which implies the time series pair has some causal structure) and is likely greater than zero (which implies the causal inference made with the leaning agrees with intuition).  These conclusions, however, depend on whether or not the t-test is applicable to this example.  For example, the assumption that the sample mean of the leanings can be assumed to follow a normal distribution based on the central limit theorem \cite{Casella2002} may rely on the sampled time series from which the leaning were calculated being independent and identically distributed, which may not be true.  The assumptions used in these statistical tests are intended to be illustrative.  Such assumptions should be explored in depth to formally develop a statistical test for causal inference using leanings.  The sampling procedure used in this example may not be applicable to other data sets for which the leaning may still be a useful causal inference tool.  Thus, it may not be possible that a single statistical test will be appropriate for all sets of leaning calculations.  

A bootstrapping \cite{Efron1994} procedure can be set up with the sample of leaning calculations, whereby $10^6$ means are calculated from new sets (of the same size as the original set) of leanings that have been sampled (with replacement) from the original set.  This procedure yields no negative means; the null hypothesis that the mean leaning value is actually negative (i.e., $\langle\lambda_1^z\rangle < 0$) can be rejected with a $p$-value less than $10^{-6}$.  The $90\%$ confidence interval for the mean of the $10^6$ bootstrapped means is $[3.48\times 10^{-3},3.57\times 10^{-3}]$, which, again, implies the mean leaning for this example is positive.    

\section{Spurious Leanings}
Consider the linear system of
\begin{eqnarray}
\label{eqn:Spur}
\left\{\mathbf{X},\mathbf{Y}\right\} = \left\{\{x_t\},\{y_t\}\right\}
\end{eqnarray}
where $t=0,1,2,\ldots,L$,
\begin{equation*}
x_t = \left\{
  \begin{array}{lr}
    2 & t = 1\\
    0 & \forall\; t\in\{t\;|\;t\neq 1 \;\mathrm{and}\; t\bmod 5 \neq 0\}\\
    2 & \forall\; t\in\{t\;|\;t\bmod 5 = 0\}
  \end{array}
\right.
\end{equation*}
and
\begin{equation*}
y_t = \eta_t
\end{equation*}
with $\eta_t\sim\mathcal{N}\left(0,1\right)$.  The first time series, $\mathbf{X}$, is the periodic impulse that drove the example system in Eqn.\ \ref{eqn:IReqn}.  The second time series, $\mathbf{Y}$, is standard Gaussian noise applied at each time step.  

\begin{figure}[ht]
\includegraphics[scale=0.45]{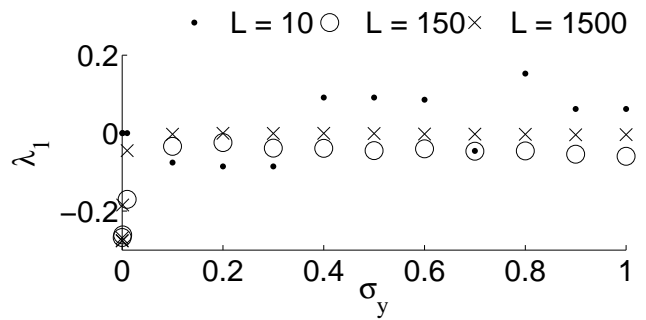}
\caption{Eqn.\ \ref{eqn:Spur} leads to spurious leanings, i.e.\ weighted mean observed leanings using the 1-standard assignment, $\lambda_1$, that depend on both the tolerance domain $\sigma_y$ and the library length $L$.}
\label{fig:nocause}
\end{figure}

There is no intuitive causal relationship in Eqn.\ \ref{eqn:Spur}.  However, Figure \ref{fig:nocause} shows the weighted mean observed leaning using the 1-standard assignment may lead to spurious causal inferences for different symmetric tolerance domains $\delta_y$, given $\delta_x = 0$.  The causal inference becomes inconclusive as the library length $L$ is increased; i.e.,\ the leaning moves towards zero for the tested tolerance domains as the library length of Eqn.\ \ref{eqn:Spur} is increased.  However, the use of leanings for causal inference with Eqn.\ \ref{eqn:Spur} at smaller library length, e.g., $L=10$, may imply a spurious relationship. 

The spurious leanings shown in this example may imply causal relationships that do not exist in the system.  Leaning calculations may be part of an exploratory causal inference analysis, but care must to be taken to ensure a causal relationship is actually present in the data, even if the directionality (or other features) of that relationship are unknown.  The relationship between leanings and causality as it is typically understood in physics (i.e., involving interventions into the systems under investigation, e.g., through experiments \cite{Pearl2000}) is not currently known.  This article is exploring the use of leanings as part of an exploratory causal analysis in times series data, not as a definition or proof of causality in a dynamical system.   

\section{Conclusion}
Causal inference using observational data alone is a difficult task \cite{kleinberg2012}.  This problem is important in many fields, but in physics in particular, there are often subfields for which direct experimentation is not technologically feasible. 

Exploratory causal analysis, as it has been described here, involves many different techniques, including Granger causality (GC), transfer entropy (TE), cross correlation (CC), and state space reconstruction (SSR).  Each of these techniques has well-known shortcomings.  For example, GC is parametric, TE can be computationally difficult \cite{kaiser2002}, CC can be unreliable \cite{Rogosa1980}, and SSR relies on correctly setting lag times and embedding dimensions \cite{Small2004}.  Causal leaning has been introduced to overcome many of these shortcomings: it is non-parametric, may be based on counting, and the only adjustable parameters are the tolerance domains and cause-effect assignments.  It can be shown that a Granger causality statistic and transfer entropy imply causal inferences that agree with intuition for examples similar to those shown in Section \ref{sec:IR}, \ref{sec:rlcirc}, \ref{sec:nonli}, and \ref{sec:2Pop}.  Thus, the leaning is not unique in its ability to provide intuitive causal inferences for these simple examples, despite a different operational definition of causality than either of these tools.  It can also be shown that a Granger causality statistic can provide the counter-intuitive causal inference for the snowfall example shown in Section \ref{sec:emp}, for which the leaning provides the intuitive causal inference.  Understanding the differences and similarities between different time series causality measures (with different operational definitions of causality) may help provide a deeper understanding of how time series causality relates (in general) to fundamental or intuitive notions of causality.  These ideas will be explored in future work.

No attempt has been made to interpret causal leanings in terms of current philosophical causality studies.   For example, there is no exploration of how causal leanings are associated with token or {\em prima facie} causality \cite{kleinberg2012}.  We have grouped the leaning method under the broad term of time series causality inference, which implies the technique is distinct from other data causality methods, including direct acyclic graph (DAG) \cite{Pearl2000} and temporal logic \cite{kleinberg2012} techniques.  Causal leanings have been introduced here as a practical tool and connections with the broader fields of data causality and causality foundations are left for future work.  For example, leanings may be a subset of the more general temporal logic presented by Kleinberg \cite{kleinberg2012} and may have interpretations within Good's probabilistic causal framework of propensities and weights of evidence \cite{Good1984}.

There are many open questions regarding the use of leanings for causal inference that have not been explored in this article.  For example, how should the magnitude of the leaning calculations be interpreted?; if there are two weighted mean observed leanings $\lambda_1$ and $\lambda_2$ with different cause-effect assignments such that $\lambda_1 > \lambda_2$ -- does the first cause-effect assignment represent a ``stronger'' driver than the second?; how should leanings of $0$, $2$, or $-2$ be interpreted with respect to the cause-effect assignments?; and how should leanings calculated using different cause-effect assignments be compared?

There has also been no formal exploration of using leanings as part of statistical tests, as is often done with GC \cite{Pierce1977}.  The use of histograms in Figure \ref{fig:dsthist} may be considered a first step toward statistical interpretations of leanings.

Finally, this article has discussed the use of leanings as part of an exploratory causal analysis of time series data.  Exactly how such an analysis should be conducted is, however, still an open question.  For example, given a report of GC, TE, SSR, and leanings (all calculated in various ways), how should the results be interpreted holistically?  There are many potentially confusing scenarios in which, e.g., two techniques lead to opposite causal inferences.  The most reasonable time series causality techniques to use for a given exploratory causal analysis may depend strongly on the data itself, but general guidelines for such analysis is, as far as we know, unknown.

\bibliography{main}

\end{document}